\documentclass[%
 reprint,
 amsmath,amssymb,
 aps,
floatfix,
superscriptaddress, 
]{revtex4-1}

\usepackage{appendix}
\usepackage{graphicx}
\usepackage{dcolumn}
\usepackage{bm}
\usepackage{xcolor}
\usepackage{float}
\usepackage{amsmath}
\usepackage{verbatim}

\usepackage{siunitx}
\usepackage[colorlinks=true, allcolors=blue]{hyperref}
\usepackage[normalem]{ulem}
\usepackage{listings}
\definecolor{mygreen}{rgb}{0,0.6,0}
\definecolor{mygray}{rgb}{0.5,0.5,0.5}
\definecolor{mymauve}{rgb}{0.58,0,0.82}

\renewcommand{\Re}{\mathfrak{Re}}

\begin{document}

\title{Light-based Chromatic Aberration Correction of Ultrafast Electron Microscopes}

\author{Marius Constantin Chirita Mihaila}
\thanks{MCCM and NLS with equal contribution. Contact authors   : marius.chirita@matfyz.cuni.cz, neli.streshkova@matfyz.cuni.cz}

  \affiliation{Charles University, Faculty of Mathematics and Physics, Ke Karlovu 3, 121 16 Prague 2}

 \author{Neli Laštovičková Streshkova}
\thanks{MCCM and NLS with equal contribution. Contact authors   : marius.chirita@matfyz.cuni.cz, neli.streshkova@matfyz.cuni.cz}

  \affiliation{Charles University, Faculty of Mathematics and Physics, Ke Karlovu 3, 121 16 Prague 2}

\author{Martin Kozák}
  \affiliation{Charles University, Faculty of Mathematics and Physics, Ke Karlovu 3, 121 16 Prague 2}

\date{\today}

\begin{abstract}

We propose and theoretically demonstrate a technique that allows one to compensate for chromatic
aberrations of traditional electron lenses in ultrafast electron microscopes. The technique is based
on space- and time-dependent phase modulation of a pulsed electron beam using interaction with
a shaped pulsed ponderomotive lens. The energy-selective focal distance is reached by combining
the electron temporal chirp with the time-dependent size of the effective potential, with which the
electrons interact. As a result, chromatic aberration can be reduced by up to a factor of seven. This
approach paves the way for advanced transverse and longitudinal wavefront shaping of electrons in
free space.
\end{abstract}

\maketitle

\section{\label{sec:level1}Introduction}
Spatial resolution of imaging using photons or electrons is fundamentally limited by the wavelength of the substance used to carry the information from the sample to the detector. However, reaching the diffraction limit of resolution requires an aberration-free imaging system. The challenge of achieving atomic resolution in low-energy electron microscopy comes mainly from aberrations in electron optics. Especially chromatic aberrations severely deteriorate the imaging performance at low electron energies ~\cite{weissbacker2001electrostatic,kaiser2011transmission,linck2016chromatic}.

The foundation for understanding and compensating for these aberrations was laid in 1936~\cite{scherzer1936einige,rose2009historical}. The Scherzer theorem established that chromatic and spherical aberrations are inevitable in rotationally symmetric electron lenses. This theorem underscored a critical limitation in electron microscopy, setting the stage for decades of research aimed at overcoming these inherent aberrations.

Later, the correction of spherical aberrations using hexapole correctors was experimentally demonstrated~\cite{zach1995aberration,haider1998spherical}. This breakthrough not only proved the theoretical proposals suggested earlier but also achieved a substantial improvement in resolution, making a significant leap in electron microscopy capabilities. 

Ultrafast electron microscopes provide exceptional temporal and spatial resolution ~\cite{zewail2010four,morimoto2018diffraction, morimoto2023attosecond}. Future progress in this field is anticipated through the integration of highly coherent field emission sources~\cite{feist2017ultrafast, houdellier2018development, zhu2020development}, aberration-corrected probes, and increased probe current.

Although programmable and adaptive optics, such as spatial light modulators (SLM), have revolutionized light optics~\cite{maurer2011spatial}, the development of programmable and adaptive phase plates for electron optics remains in its early stages~\cite{grillo2014,SHILOH201426,shiloh2018spherical,VERBEECK201858, ibanez2022,roitman2021shaping, ribet2023design, yu2023quantum}. Recent research has explored the coherent manipulation of electron waves through interactions with optical near-fields~\cite{barwick2009photon,feist2015quantum,vanacore2018attosecond,vanacore2019ultrafast,konevcna2020electron,ben2021shaping,shiloh2021electron,henke2021integrated,dahan2021imprinting,feist2022cavity,madan2022ultrafast,tsesses2023tunable,garcia2023spatiotemporal,gaida2023lorentz,synanidis2024quantum,fang2024structured} and pure optical fields~\cite{Bucksbaum1988,freimund2001observation,freimund2002bragg,dwyer2006femtosecond,baum2007attosecond, Kozak2018, schwartz2019laser, axelrod2020observation, chirita2022transverse, tsarev2023nonlinear, ebel2023inelastic,schaap2023ponderomotive,velasco2024free}  in free space. Furthermore, theoretical proposals have demonstrated the correction of spherical aberrations in electron lenses via controlled light interactions~\cite{de2021optical, uesugi2021electron, uesugi2022properties, ChiritaMihaila:25}.

We theoretically demonstrate here a method to compensate for chromatic aberrations in traditional electron lenses in ultrafast electron microscopes by pairing their positive chromatic aberration with a ponderomotive lens exhibiting negative chromatic aberration. Independent of our work, a recent study~\cite{nekula2025laser} also investigated this effect using ray-tracing methods and reported results consistent with ours. In contrast, our approach is based on a wave-optics model, providing a complementary and rigorous perspective. The correction method proposed here is highly effective at low electron energies; but at energies above 5 keV, the improvement of the chromatic coefficient value and the associated electron beam profile in focus becomes less pronounced.

Unlike multipole correctors~\cite{linck2016chromatic}, this method achieves the correction in a single interaction plane, potentially reducing the complexity, spatial requirements, and costs of aberration correctors. Using the interaction between spatially shaped pulsed light and a chirped electron pulse, specifically utilizing the phase shift gradient encountered by the electron pulse at moderate convergence angles of the pulsed optical lens, this approach offers a promising solution to a long-standing challenge in electron microscopy.

\section{\label{sec:level2}Theoretical considerations}

The aberration function of electrostatic and magnetostatic electron lenses, which accounts for the chromatic aberration, characterizes the phase shift distortion of the electron wave and can be expressed as~\cite{haider2000upper}:
\begin{equation}
    \chi(\alpha,E)\approx  \frac{\pi}{\lambda_e(E)}\Big[\Delta z_0 +C_c\frac{E-E_0}{E_0}\Big]\alpha^2.
    \label{eqn: Aberration function trad lens}
\end{equation}

\begin{figure*}
   \centering      
   \includegraphics[width=17.0cm]{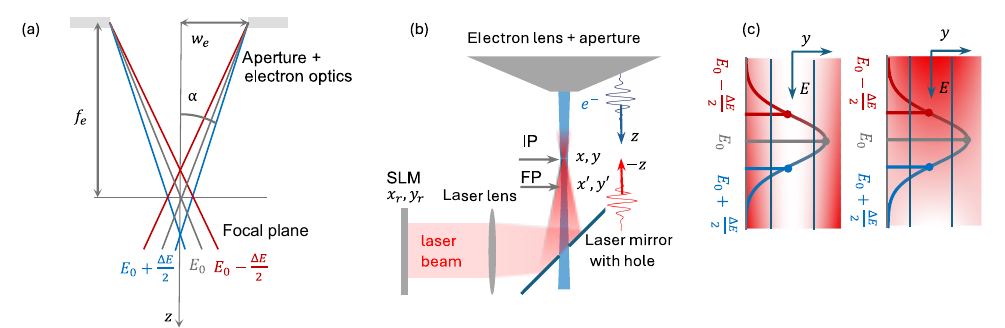} \caption{(a) Electron trajectories with positive chromatic aberration. Due to the energy spread of the electron beam, slower electrons will get focused before the geometric focal plane and faster electrons after. (b) Scheme of the experimental geometry. The optical beam is shaped in front of the SLM plane ($x_{r},y_{r}$), and focused near the interaction plane IP ($x,y$) where the chromatic aberration of the electrons is corrected. Subsequently, the electrons are focused in the Fourier plane (FP) ($x',y'$). (c) Interaction with a ponderomotive lens generated by a vortex-like (left) and Gaussian-like beam (right). The red color scale illustrates a snapshot of the longitudinal intensity cross-section, $g^2(y,z)$. The interaction plane, with a radius of $<\SI{2}{\micro m}$, where the shaped laser intensity is used for chromatic correction, is shown within the dashed blue lines. Furthermore, a snapshot of the electron beam is represented in it's rest frame, where electrons with  energy $E_0+\Delta E/2$ are placed forward and electrons with energy $E_0 - \Delta E/2$ are placed backward. The vortex-like beam causes the high energy parts of the electron beam to converge more strongly than the low energy parts. The Gaussian-like beam causes the low energy parts to diverge more strongly than the high energy part. Both induce negative chromatic aberration. While it is possible to irradiate the electron beam from different directions, we considered laser irradiation from below primarily to avoid practical issues such as unwanted illumination of the electron detector.}
   \label{fig:chromatic aberration sketch}
\end{figure*}

\noindent with $\alpha$ the convergence semi-angle, the modified defocus due to the electron energy spread $\Delta z_0 + C_c\frac{E-E_0}{E_0}$, where $C_c$ is the chromatic aberration coefficient and $\Delta z_0$ is the defocus that can be tuned by the electron lens. Furthermore,  $E_0$  is the mean electron energy and $E$ is the actual electron energy within the total energy spread $\Delta E$ of the electron beam. The energy-dependent relativistic electron wavelength is defined as $\lambda_e(E) = 2\pi \hbar c/(\sqrt{2EE_{r_0}+E^2})$, with the rest electron energy $E_{r_0} = m_e c^2$, $\hbar$ the reduced Planck constant, $m_e$ the electron rest mass, and $c$ the light velocity in vacuum. For brevity, we denote $\chi_0 = \chi(\Delta z_0 = 0)$ solely the energy-dependent part of the aberration function.

Due to the energy spread of the electrons, faster electrons passing through traditional lenses get focused after the geometrical focal plane, whereas slower electrons focus before it (see Fig.~\ref{fig:chromatic aberration sketch}(a)). 

In the particle picture, when the electrons interact with strong optical fields in vacuum, transverse momentum from the photons is transferred to the electrons, influencing their trajectory. The electrons are repelled from areas with high photon density~\cite{kibble1966mutual}. A suitably shaped optical field can thus act as a lens for the electron beam. To achieve the chromatic correction it is necessary to generate the light lens with different tailored shapes for the different energy components within the electron beam. This can be accomplished by using a focused optical beam propagating against the electron beam  (see Fig.~\ref{fig:chromatic aberration sketch}(b)). Due to the pulsed electron beam, which has defined energy-time correlation (chirp), different energy components arrive to the site of interaction with the light field at different times. With proper setting of the synchronization of the electron and the optical pulse out of the optical beam focus, different energy components of the electron pulse experience different lensing strength (see Fig.~\ref{fig:chromatic aberration sketch}(c)).

The ponderomotive chromatic aberration correction creates a linearly increasing convergent lens effect, where faster electrons experience stronger lensing than slower ones (see Fig.~\ref{fig:chromatic aberration sketch}(c) left). Conversely, when using divergent lensing for chromatic correction, slower electrons should experience stronger lensing, which decreases linearly for faster electrons (see Fig.~\ref{fig:chromatic aberration sketch}(c) right).

During the interaction with a laser field the electron wave will acquire a ponderomotive phase shift, which is defined as~\cite{de2021optical,chirita2022transverse}:
\begin{equation}
    \varphi(\mathbf r, t) = -(1/\hbar) \int_{-\infty}^\infty \mathrm{d} t \, U (\mathbf r,t),
    \label{eq:phase_original}
\end{equation}
with the ponderomotive potential:
\begin{equation}
U (\mathbf r,t) = \frac{e^2}{2m_e \gamma } \left[ A_x^2(\mathbf r, t) + A_y^2(\mathbf r, t) + \frac{A_z^2(\mathbf r, t)}{\gamma^2} \right],
\label{eq:Ponderomotive pontential original}
\end{equation}
where $\gamma = 1/\sqrt{1-v_e^2/c^2}$ is the relativistic factor, $v_e = c \sqrt{1 -1/\left(1 + \frac{E}{E_{r_{0}}}\right)^2} $ is the relativistic electron velocity in the $z$ direction and $\mathbf{A(\mathbf{r},t)}$ is the vector potential. We consider a harmonic optical pulse traveling in the $-z$ direction:
 
\begin{equation}
    \mathbf{\varepsilon}(\mathbf{r},t) = \mathbf{\Tilde{g}}(\mathbf{r}) \exp\left[-\frac{(t - t_0 + \frac{z}{c})^2}{2\sigma_t^2}\right]e^{-i \omega  (t + \frac{z}{c}) },
\end{equation}
where $\mathbf{\Tilde{g}}(\mathbf{r})$ is the spatial profile of the laser field (see Supplemental Material for details). The time envelope is described as a Gaussian pulse with \( t_0 \) the time at the center of the pulse, \( \frac{z}{c} \) accounting for the propagation along the negative \( z \)-axis,  \( \sigma_t \) the temporal duration of the pulse, and \(\omega\) the angular frequency. Considering that the envelopes vary slowly compared to the carrier frequency, we can use the simplification $\mathbf{A}(\mathbf{r}, t) =\Re{\Big[\frac{\bm{\varepsilon}(\mathbf{r}, t)}{i\omega}\Big]}$~\cite{chirita2022transverse}. The ponderomotive phase is then:
\begin{equation}
    \begin{split}
        \label{eq:Phase shift to be calculated}
        \varphi(\mathbf{r},t) = & -\frac{e^{2}}{2m_{e}\gamma\hbar\omega^{2}} \int \mathrm{d}t\,  \Big[ |\Tilde g_{x}|^{2}(\mathbf{r})+|\Tilde g_{y}|^{2}(\mathbf{r})+\frac{1}{\gamma^{2}}|\Tilde g_{z}|^{2}(\mathbf{r}) \Big] \\ & \times \exp\left[-\frac{\left(t - t_0 + \frac{z}{c}\right)^2}{\sigma_t^2}\right] \sin^2{[ \omega (t + z/c)]}.
    \end{split}
\end{equation}
When using a focused optical beam the ponderomotive phase $\varphi(\mathbf{r},t)$ has a gradient along the $z$ axis. The coupling between the chirped electron pulse energy components $E$ and the graded ponderomotive phase coordinate $z$ occurs thanks to the electron pulse chirp and also the energy dependent velocity of the electrons. To calculate the ponderomotive phase in the rest frame of each energy component $z'(E)$ we substitute for the laboratory frame coordinate $ z(E) = z_0'(E) + v_e t$, where the first term models the longitudinal electron energy distribution due to the chirp  $z_0'(E) =  (E - E_0) \tau_e  v_0 / 2 \Delta E$ and the second term $v_e t$ accounts for the different velocities. In the linearized chirp model, $\tau_e$ is the duration of the electron pulse and $v_0$ is the electron velocity at $E_0$. Further, we can write $\sin^{2}{[\omega(t+z/c)]=(1-\cos[\omega(t+z/c)})/2]$ and neglect the integral over the cosine term due to the slowly varying envelope. Finally, the energy dependent ponderomotive phase is:

\begin{equation}
    \begin{split}
        \label{eq:Phase shift E}
        \varphi(x, y, E) \approx & -\frac{e^{2}}{4m_{e}\gamma\hbar\omega^{2}} \int \mathrm{d} t\,  \Big[ 
        |\Tilde{g}_{x}|^{2}(x,y,E)+|\Tilde{g}_{y}|^{2}(x,y,E) \\ & +\frac{1}{\gamma^{2}}|\Tilde{g}_{z}|^{2}(x,y,E) \Big] \\ & \times \exp\left[-\frac{\left(t - t_0 + \frac{z_0'(E)+v_e t}{c}\right)^2}{\sigma_t^2}\right].
    \end{split}
\end{equation}
We further employ the Bluestein algorithm ~\cite{Yanlei2020}, to enable the propagation and shaping of an initial Gaussian optical field into the desired mode $\mathbf{\Tilde{g}}(\mathbf{r})$ used for the interaction with electrons (see Supplemental Material). This propagation method is faster than other propagation methods and is less sensitive to sampling, making it particularly suitable for this application. A successful aberration correction is achieved when the concavity of the sum of the aberration phase and the ponderomotive phase $\chi_0(x,y, E) + \varphi(x,y, E)$ does not depend on the energy of the electrons $E$, making all of the energy components to focus at the same plane. This condition corresponds to minimizing $\frac{\partial}{\partial E}\frac{\partial^{2}}{\partial \alpha^{2}}[\chi_0(x, y, E) + \varphi(x,  y, E)] $. 

If we denote the initial wavefunction as $\psi_{0}$ in the electron-light interaction plane, the intensity profile of the electron beam in the focal plane can be calculated as the incoherent sum of the contributions of the different energies within the beam~\cite{haider2000upper}:
\begin{align}
    I(x',y') &= \int_{-\infty}^{\infty} \mathrm{d}E\, |\psi(x',y',E)|^{2} 
     \notag \\
    &\quad \frac{1}{\sqrt{2\pi\sigma^2_E}} \times \exp{\Bigg[ 
    -\frac{(E-E_{0})^{2}}{2\sigma^2_E}
    \Bigg]},
    \label{Electron probe intensity}
\end{align}
where $\psi(x',y',E) =  \mathcal{F}_{2D}\left(e^{i\chi(x,y,E)}e^{i\varphi(x,y,E)}\psi_{0}(x,y,E)\right)$, $\mathcal{F}_{2D}$ is the 2D Fourier transform and $\sigma^2_E = \Delta E^2/(8\ln{2})$ is the standard deviation squared of the energy distribution.

\section{\label{sec:level3}Results}
We use Zernike polynomials~\cite{lakshminarayanan2011zernike} (see Supplemental Material) to precisely tailor the shape of the ponderomotive lenses. We denote the modified laser beams further as vortex-like and Gaussian-like, referring, respectively, to the modified vortex and Gaussian beam.  We carry out the calculations and visualize the phases in 2D in the $yE$ -plane, considering only the middle slice of the phase distribution.
\begin{figure}
    \centering
    \includegraphics[width=8.0cm]{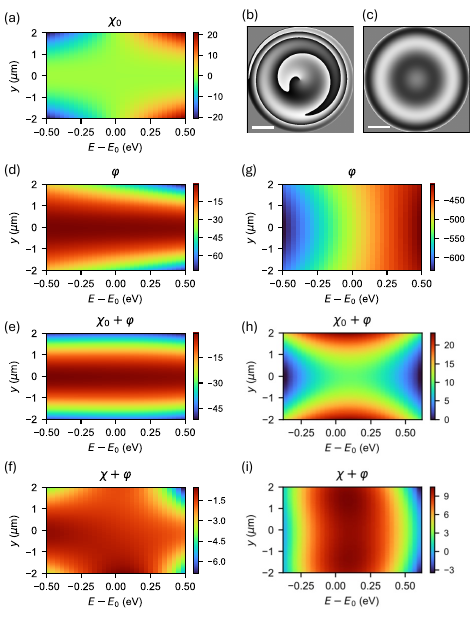}
    \caption{ Correction of chromatic aberration in traditional electron lenses and the compensation of ponderomotive phase shifts. The electron phase shifts, expressed in radians, are computed for each electron energy E. However, to enhance visualization, the x-axis displays $E-E_0$ instead. (a) Energy-dependent chromatic aberration phase shift of the electron lens. (b) Vortex-like phase profile (scale bar: $\SI{1}{cm}$; phase ranges from $-\pi$ to $\pi$, white to black) and (c) Gaussian-like phase profile used to shape the focal spots shown in Figs.~\ref{fig:S1}~(b) and (d), respectively.
 (d) Ponderomotive phase shift when using a shaped vortex laser beam. (e) Combined aberration phase and ponderomotive phase shift without and (f) with electron lens defocus. (g) Energy dependent ponderomotive phase shift when using a Gaussian-like beam. (h) Combined aberration phase and ponderomotive phase shift for a Gaussian-like beam without and (i) with electron lens defocus. }
    \label{fig:phases}
\end{figure}

The energy-dependent part of the aberration phase $\chi_0$ at $x=0$ as a function of $y$ and $E$ is shown in Fig. \ref{fig:phases}(a). At low electron energies, of the order $\SI{1}{keV}$, the contribution of chromatic aberration becomes dominant compared to other aberrations, significantly affecting the size of the electron probe and limiting its resolution~\cite{haider1998spherical}. As such, we model the electron beam with the chromatic aberration coefficient $C_c=8\,\rm{mm}$, central energy of $E_0 = \SI{1}{keV}$, FWHM energy spread $\Delta E=0.5\,\rm{eV}$, and a maximum electron convergence angle of $\alpha_{\text{max}}=8\, \rm{mrad}$. The interaction with light is placed close to the electron cross-over, where the electron beam radius is $w_e=\SI{2}{\micro m}$.

The chromatic aberration of the electron lens is corrected using either an optical vortex-like or a Gaussian-like beam, with a tailored phase and intensity profile, imprinted by an SLM positioned at the point where the optical beam is collimated. A Gaussian-like beam, with a central wavelength of $\lambda_0 = 2060\,\mathrm{nm}$, a pulse duration of $\tau_0 = 250\,\mathrm{fs}$, and focused using a numerical aperture ($\mathrm{NA}$) of $0.14$, requires approximately $\SI{4}{\micro J}$ of pulse energy for chromatic correction. In comparison, a vortex-like beam demands about $\SI{16}{\micro J}$ of pulse energy for the same application. The duration of the electron pulse is $\tau_e = 1200 \,\rm{fs}$ and the central energy group velocity is $v_0=0.06c$. The temporal overlap of the electron pulse and the optical pulse envelopes is defined at the optical beam focus ($z=0$) when the coordinate systems of the electron and laser beams are aligned without spatial or temporal shifts. The shift in the optical beam focus relative to the temporal overlap is controlled by applying a lens phase to the optical fields. Figure~\ref{fig:phases}(b) shows the Zernike-optimized phase profile applied to the incoming Gaussian beam to achieve the intensity distribution depicted in the top panel of Fig.\ref{fig:chromatic aberration sketch}(c), while Fig.\ref{fig:phases}(c) illustrates the phase profile used to produce the intensity distribution shown in the bottom panel of Fig.~\ref{fig:chromatic aberration sketch}(c). 

The vortex-like beam generates a ponderomotive phase shift, as shown in Fig.~\ref{fig:phases}(d), which acts as a convergent lens. This effect causes stronger focusing on the higher-energy electrons, shifting their focal point backward. The final phase profile of the electron beam, which is given as $\chi_0(x,y, E) + \varphi(x,y, E)$ has weaker energy dependence, as illustrated in Fig.~\ref{fig:phases}(e). Adding an energy-independent positive defocus of $\Delta z_0 = 8.5\, \rm{\mu m}$ can bring the focal point back to the geometrical focus of the $E_0$ electron beam component, resulting in a nearly constant phase (see Fig.~\ref{fig:phases}(f)). The slight asymmetry in Fig.~\ref{fig:phases}(f) is caused by the finite resolution of the SLM resulting in the phase in Fig.~\ref{fig:phases}(b), causing an effective axial misalignment of the focused optical beam. 

The correction with a Gaussian-like mode is analogous, with the ponderomotive phase shift in Fig.~\ref{fig:phases}(g) acting as an energy-dependent divergent lens in the energy difference region between -0.5 eV and 0.25 eV and pushing the less energetic components of the electron beam forward  (see Fig.~\ref{fig:phases}(h)). Then, as a result of the shaping, the ponderomotive potential curvature flips, and the ponderomotive phase acts as a convergent lens. Negative defocus of $\Delta z_0 = -2.5 \, \rm{\mu m}$ results in phase constant in the $y$ direction (see Fig.~\ref{fig:phases}(i)). In Figs. \ref{fig:phases}(h,i) we subtracted the $y$-independent component of the phase, which is linear in $E$. This component of the phase has no effect on the electron focal spot and only induces constant longitudinal acceleration, which can be neglected in this case. The linear phase change is $\approx200\,\rm{rad}$, which we estimate to yield an energy shift of $\approx 0.12\, \rm{eV}$. Note that the acceleration can be tuned by shaping the longitudinal intensity profile of the laser beam, enabling advanced longitudinal wavefront shaping of electrons. 

\begin{figure}
    \centering
    \includegraphics[width=\linewidth]{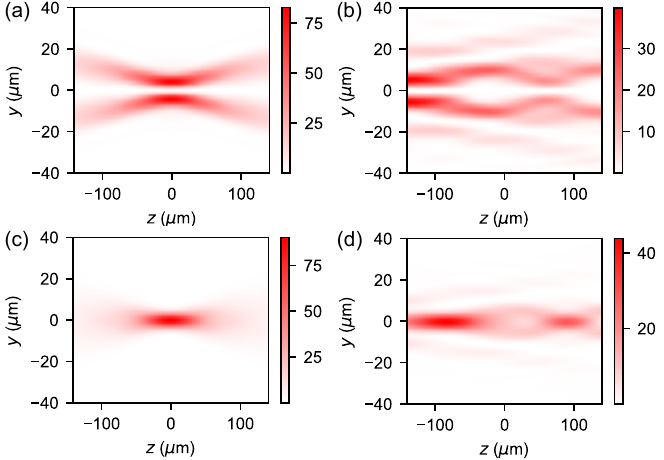}
    \caption{(a) Focus of a simple Laguerre-Gauss beam. (b) Focus of phase-shaped vortex-like beam used for chromatic correction. (c)  Focus of a simple Gaussian beam. (d) Focus of a phase-shaped Gaussian-like beam used for chromatic correction.
}
    \label{fig:S1}
\end{figure}

Figures~\ref{fig:S1}~(a) and (c) are the unshaped Laguerre-Gauss and Gauss beams.  Figures~\ref{fig:S1} (b) and (d) show the focal planes of the vortex-like and Gauss-like beams, respectively, which are used for chromatic aberration correction.

The ponderomotive phase $\varphi$ is approximately proportional to the intensity profile and the interaction region is delimitated by $\approx \pm10\, \rm{\mu m}$ in the $z$ direction and $\pm 2\,\rm{\mu m}$ in the $y$ direction. We note that the intensity profiles shown in Fig.~\ref{fig:S1} (b) and (d) undergo several rounds of focusing and defocusing, maintaining the convergent lens properties for the vortex-like beam.  In contrast, the intensity distribution of the focused Gaussian-like beam changes from concave to convex around $z=0$, also changing the lensing from divergent to convergent.

The electron beam focal spot with the chromatic aberration  $C_c=8\,\rm{mm} $ is shown in Fig. \ref{fig:focii}(a). The resulting electron profile after the ponderomotive phase shift is shown in Fig.~\ref{fig:focii}(b) for a vortex-like beam and is similar for a Gaussian-like beam (not shown here). We note that the intensity of the central peak after both corrections is over $80\%$ of the target peak intensity compared to $25\%$ in the aberrated beam intensity and that the energy dispersed in the background is significantly suppressed. Through correlation with beam focus profiles for various $C_c$ we estimate that the effective $C_{c,\text{eff}}$ achieved through correction is around $1.2\,\rm{mm}$ for the vortex-like beam and $1.1\,\rm{mm}$ for the Gaussian-like beam, resulting in improvement by a factor of $7$ for these parameters. Other criteria demonstrating chromatic correction include the reduction in the standard deviation of the electron probe profiles: $\sigma_{\text{vortex}} = 2.67\,\mathrm{nm}$ and $\sigma_{\text{gauss}} = 2.72\,\mathrm{nm}$, compared to the aberrated beam $\sigma_{\text{aber}} = 6.07\,\mathrm{nm}$ and the target beam $\sigma_{\text{goal}} = 2.39\,\mathrm{nm}$. Furthermore, there is a clear improvement in the contrast between the central peak and the background intensity.

\section{\label{sec:level4}Discussion and Conclusion}
\begin{figure}
    \centering
    \includegraphics[width=\linewidth]{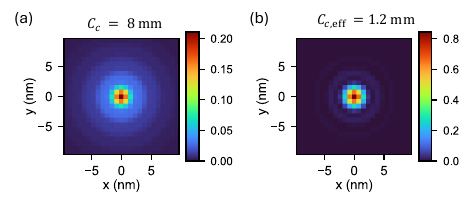}
    \caption{(a) Focus of a simple vortex beam. (b) Electron focal spot after correction with a vortex-like beam. The simulations are carried out using Eq.~\ref{Electron probe intensity}.}
    \label{fig:focii}
\end{figure}

Our study introduces a novel approach to mitigate chromatic aberrations in ultrafast electron microscopes using the ponderomotive potential generated by shaped optical fields. This method effectively compensates for chromatic distortions by tailoring the lensing strength to different energy components of a chirped electron beam. By creating a phase gradient along the longitudinal axis of the electron pulse, this strategy achieves energy-dependent ponderomotive lensing, reducing chromatic aberrations by approximately a factor of seven and significantly enhancing spatial resolution. In a Cc-uncorrected microscope, obtaining an equivalent focus spread (see Fig.~\ref{fig:focii} (b)) with a monochromator~\cite{kaiser2011transmission,kimoto2014practical,streshkova2024monochromatization} would require narrowing the electron gun energy distribution to $\Delta E \sim \SI{70}{meV} $.

While electron pulses can be compressed to durations with negligible chirp at specific locations along the beam path, such compression inherently implies the presence of a chirp upstream. Our method targets the pulse at a stage where it is still chirped, either prior to or during compression, making it broadly compatible with any pulsed electron source, independent of the final pulse duration at the sample.

The shaping of Gaussian and vortex beams using Zernike polynomials allowed for precise control over the light intensity distribution. This approach ensured a nearly linear variation in the axial direction and a quadratic dependence in the radial direction, optimizing the interaction with the electron pulse. In this study, phase-only shaping techniques were used and optimized using a modified gradient descent algorithm~\cite{xie2015phase}. Incorporating also amplitude modulation alongside phase shaping could further enhance chromatic correction, potentially improving the overall performance and flexibility of the approach. Furthermore, improved beam-pointing stability can be achieved by integrating active piezo-driven beam-pointing stabilization systems, utilizing optical cavities~\cite{peccianti2012demonstration}, or spatial filtering.

For experimental validation, measuring electron defocus under varying electron energy distributions~\cite{haider1998spherical}, the practical implementation of the method can be rigorously tested. Beyond chromatic aberration correction, the proposed technology may also prove valuable for applications such as electron-atom scattering using ultrashort and tightly focused electron beams, where precise control over wavefronts is essential~\cite{morimoto2024scattering}.

With ongoing advances in artificial intelligence, dynamic wavefront shaping of electrons using light may soon offer a promising solution to chromatic aberrations in ultrafast electron microscopy.


\begin{acknowledgments}
The authors acknowledge funding from the Czech Science Foundation (project 22-13001K), Charles University (SVV-2023-260720, PRIMUS/19/SCI/05, GAUK 90424) and the European Union (ERC, eWaveShaper, 101039339). Views and opinions expressed are however those of the author(s) only and do not necessarily reflect those of the European Union or the European Research Council Executive Agency. Neither the European Union nor the granting authority can be held responsible
for them. This work was supported by TERAFIT project No. \text{CZ}.02.01.01/00/22\_008/0004594 funded by OP JAK, call Excellent Research.

\noindent\textbf{Data Availability.} The data supporting the findings of this study are openly available at~\cite{zenodo_dataset}.

 \end{acknowledgments}
\bibliography{literatur}

\providecommand{\noopsort}[1]{}\providecommand{\singleletter}[1]{#1}
\begin{thebibliography}{66}%
\makeatletter
\providecommand \@ifxundefined [1]{%
 \@ifx{#1\undefined}
}%
\providecommand \@ifnum [1]{%
 \ifnum #1\expandafter \@firstoftwo
 \else \expandafter \@secondoftwo
 \fi
}%
\providecommand \@ifx [1]{%
 \ifx #1\expandafter \@firstoftwo
 \else \expandafter \@secondoftwo
 \fi
}%
\providecommand \natexlab [1]{#1}%
\providecommand \enquote  [1]{``#1''}%
\providecommand \bibnamefont  [1]{#1}%
\providecommand \bibfnamefont [1]{#1}%
\providecommand \citenamefont [1]{#1}%
\providecommand \href@noop [0]{\@secondoftwo}%
\providecommand \href [0]{\begingroup \@sanitize@url \@href}%
\providecommand \@href[1]{\@@startlink{#1}\@@href}%
\providecommand \@@href[1]{\endgroup#1\@@endlink}%
\providecommand \@sanitize@url [0]{\catcode `\\12\catcode `\$12\catcode `\&12\catcode `\#12\catcode `\^12\catcode `\_12\catcode `\%12\relax}%
\providecommand \@@startlink[1]{}%
\providecommand \@@endlink[0]{}%
\providecommand \url  [0]{\begingroup\@sanitize@url \@url }%
\providecommand \@url [1]{\endgroup\@href {#1}{\urlprefix }}%
\providecommand \urlprefix  [0]{URL }%
\providecommand \Eprint [0]{\href }%
\providecommand \doibase [0]{http://dx.doi.org/}%
\providecommand \selectlanguage [0]{\@gobble}%
\providecommand \bibinfo  [0]{\@secondoftwo}%
\providecommand \bibfield  [0]{\@secondoftwo}%
\providecommand \translation [1]{[#1]}%
\providecommand \BibitemOpen [0]{}%
\providecommand \bibitemStop [0]{}%
\providecommand \bibitemNoStop [0]{.\EOS\space}%
\providecommand \EOS [0]{\spacefactor3000\relax}%
\providecommand \BibitemShut  [1]{\csname bibitem#1\endcsname}%
\let\auto@bib@innerbib\@empty
\bibitem [{\citenamefont {Wei{\ss}b{\"a}cker}\ and\ \citenamefont {Rose}(2001)}]{weissbacker2001electrostatic}%
  \BibitemOpen
  \bibfield  {author} {\bibinfo {author} {\bibfnamefont {C.}~\bibnamefont {Wei{\ss}b{\"a}cker}}\ and\ \bibinfo {author} {\bibfnamefont {H.}~\bibnamefont {Rose}},\ }\href@noop {} {\bibfield  {journal} {\bibinfo  {journal} {Journal of electron microscopy}\ }\textbf {\bibinfo {volume} {50}},\ \bibinfo {pages} {383} (\bibinfo {year} {2001})}\BibitemShut {NoStop}%
\bibitem [{\citenamefont {Kaiser}\ \emph {et~al.}(2011)\citenamefont {Kaiser}, \citenamefont {Biskupek}, \citenamefont {Meyer}, \citenamefont {Leschner}, \citenamefont {Lechner}, \citenamefont {Rose}, \citenamefont {St{\"o}ger-Pollach}, \citenamefont {Khlobystov}, \citenamefont {Hartel}, \citenamefont {Mueller} \emph {et~al.}}]{kaiser2011transmission}%
  \BibitemOpen
  \bibfield  {author} {\bibinfo {author} {\bibfnamefont {U.}~\bibnamefont {Kaiser}}, \bibinfo {author} {\bibfnamefont {J.}~\bibnamefont {Biskupek}}, \bibinfo {author} {\bibfnamefont {J.~C.}\ \bibnamefont {Meyer}}, \bibinfo {author} {\bibfnamefont {J.}~\bibnamefont {Leschner}}, \bibinfo {author} {\bibfnamefont {L.}~\bibnamefont {Lechner}}, \bibinfo {author} {\bibfnamefont {H.}~\bibnamefont {Rose}}, \bibinfo {author} {\bibfnamefont {M.}~\bibnamefont {St{\"o}ger-Pollach}}, \bibinfo {author} {\bibfnamefont {A.~N.}\ \bibnamefont {Khlobystov}}, \bibinfo {author} {\bibfnamefont {P.}~\bibnamefont {Hartel}}, \bibinfo {author} {\bibfnamefont {H.}~\bibnamefont {Mueller}},  \emph {et~al.},\ }\href@noop {} {\bibfield  {journal} {\bibinfo  {journal} {Ultramicroscopy}\ }\textbf {\bibinfo {volume} {111}},\ \bibinfo {pages} {1239} (\bibinfo {year} {2011})}\BibitemShut {NoStop}%
\bibitem [{\citenamefont {Linck}\ \emph {et~al.}(2016)\citenamefont {Linck}, \citenamefont {Hartel}, \citenamefont {Uhlemann}, \citenamefont {Kahl}, \citenamefont {M{\"u}ller}, \citenamefont {Zach}, \citenamefont {Haider}, \citenamefont {Niestadt}, \citenamefont {Bischoff}, \citenamefont {Biskupek} \emph {et~al.}}]{linck2016chromatic}%
  \BibitemOpen
  \bibfield  {author} {\bibinfo {author} {\bibfnamefont {M.}~\bibnamefont {Linck}}, \bibinfo {author} {\bibfnamefont {P.}~\bibnamefont {Hartel}}, \bibinfo {author} {\bibfnamefont {S.}~\bibnamefont {Uhlemann}}, \bibinfo {author} {\bibfnamefont {F.}~\bibnamefont {Kahl}}, \bibinfo {author} {\bibfnamefont {H.}~\bibnamefont {M{\"u}ller}}, \bibinfo {author} {\bibfnamefont {J.}~\bibnamefont {Zach}}, \bibinfo {author} {\bibfnamefont {M.}~\bibnamefont {Haider}}, \bibinfo {author} {\bibfnamefont {M.}~\bibnamefont {Niestadt}}, \bibinfo {author} {\bibfnamefont {M.}~\bibnamefont {Bischoff}}, \bibinfo {author} {\bibfnamefont {J.}~\bibnamefont {Biskupek}},  \emph {et~al.},\ }\href@noop {} {\bibfield  {journal} {\bibinfo  {journal} {Physical review letters}\ }\textbf {\bibinfo {volume} {117}},\ \bibinfo {pages} {076101} (\bibinfo {year} {2016})}\BibitemShut {NoStop}%
\bibitem [{\citenamefont {Scherzer}(1936)}]{scherzer1936einige}%
  \BibitemOpen
  \bibfield  {author} {\bibinfo {author} {\bibfnamefont {O.}~\bibnamefont {Scherzer}},\ }\href@noop {} {\bibfield  {journal} {\bibinfo  {journal} {Zeitschrift f{\"u}r Physik}\ }\textbf {\bibinfo {volume} {101}},\ \bibinfo {pages} {593} (\bibinfo {year} {1936})}\BibitemShut {NoStop}%
\bibitem [{\citenamefont {Rose}(2009)}]{rose2009historical}%
  \BibitemOpen
  \bibfield  {author} {\bibinfo {author} {\bibfnamefont {H.~H.}\ \bibnamefont {Rose}},\ }\href {https://academic.oup.com/jmicro/article-abstract/58/3/77/1989077} {\bibfield  {journal} {\bibinfo  {journal} {Journal of electron microscopy}\ }\textbf {\bibinfo {volume} {58}},\ \bibinfo {pages} {77} (\bibinfo {year} {2009})}\BibitemShut {NoStop}%
\bibitem [{\citenamefont {Zach}\ and\ \citenamefont {Haider}(1995)}]{zach1995aberration}%
  \BibitemOpen
  \bibfield  {author} {\bibinfo {author} {\bibfnamefont {J.}~\bibnamefont {Zach}}\ and\ \bibinfo {author} {\bibfnamefont {M.}~\bibnamefont {Haider}},\ }\href@noop {} {\bibfield  {journal} {\bibinfo  {journal} {Nuclear Instruments and Methods in Physics Research Section A: Accelerators, Spectrometers, Detectors and Associated Equipment}\ }\textbf {\bibinfo {volume} {363}},\ \bibinfo {pages} {316} (\bibinfo {year} {1995})}\BibitemShut {NoStop}%
\bibitem [{\citenamefont {Haider}\ \emph {et~al.}(1998)\citenamefont {Haider}, \citenamefont {Rose}, \citenamefont {Uhlemann}, \citenamefont {Schwan}, \citenamefont {Kabius},\ and\ \citenamefont {Urban}}]{haider1998spherical}%
  \BibitemOpen
  \bibfield  {author} {\bibinfo {author} {\bibfnamefont {M.}~\bibnamefont {Haider}}, \bibinfo {author} {\bibfnamefont {H.}~\bibnamefont {Rose}}, \bibinfo {author} {\bibfnamefont {S.}~\bibnamefont {Uhlemann}}, \bibinfo {author} {\bibfnamefont {E.}~\bibnamefont {Schwan}}, \bibinfo {author} {\bibfnamefont {B.}~\bibnamefont {Kabius}}, \ and\ \bibinfo {author} {\bibfnamefont {K.}~\bibnamefont {Urban}},\ }\href@noop {} {\bibfield  {journal} {\bibinfo  {journal} {Ultramicroscopy}\ }\textbf {\bibinfo {volume} {75}},\ \bibinfo {pages} {53} (\bibinfo {year} {1998})}\BibitemShut {NoStop}%
\bibitem [{\citenamefont {Zewail}(2010)}]{zewail2010four}%
  \BibitemOpen
  \bibfield  {author} {\bibinfo {author} {\bibfnamefont {A.~H.}\ \bibnamefont {Zewail}},\ }\href@noop {} {\bibfield  {journal} {\bibinfo  {journal} {science}\ }\textbf {\bibinfo {volume} {328}},\ \bibinfo {pages} {187} (\bibinfo {year} {2010})}\BibitemShut {NoStop}%
\bibitem [{\citenamefont {Morimoto}\ and\ \citenamefont {Baum}(2018)}]{morimoto2018diffraction}%
  \BibitemOpen
  \bibfield  {author} {\bibinfo {author} {\bibfnamefont {Y.}~\bibnamefont {Morimoto}}\ and\ \bibinfo {author} {\bibfnamefont {P.}~\bibnamefont {Baum}},\ }\href@noop {} {\bibfield  {journal} {\bibinfo  {journal} {Nature Physics}\ }\textbf {\bibinfo {volume} {14}},\ \bibinfo {pages} {252} (\bibinfo {year} {2018})}\BibitemShut {NoStop}%
\bibitem [{\citenamefont {Morimoto}(2023)}]{morimoto2023attosecond}%
  \BibitemOpen
  \bibfield  {author} {\bibinfo {author} {\bibfnamefont {Y.}~\bibnamefont {Morimoto}},\ }\href@noop {} {\bibfield  {journal} {\bibinfo  {journal} {Microscopy}\ }\textbf {\bibinfo {volume} {72}},\ \bibinfo {pages} {2} (\bibinfo {year} {2023})}\BibitemShut {NoStop}%
\bibitem [{\citenamefont {Feist}\ \emph {et~al.}(2017)\citenamefont {Feist}, \citenamefont {Bach}, \citenamefont {da~Silva}, \citenamefont {Danz}, \citenamefont {M{\"o}ller}, \citenamefont {Priebe}, \citenamefont {Domr{\"o}se}, \citenamefont {Gatzmann}, \citenamefont {Rost}, \citenamefont {Schauss} \emph {et~al.}}]{feist2017ultrafast}%
  \BibitemOpen
  \bibfield  {author} {\bibinfo {author} {\bibfnamefont {A.}~\bibnamefont {Feist}}, \bibinfo {author} {\bibfnamefont {N.}~\bibnamefont {Bach}}, \bibinfo {author} {\bibfnamefont {N.~R.}\ \bibnamefont {da~Silva}}, \bibinfo {author} {\bibfnamefont {T.}~\bibnamefont {Danz}}, \bibinfo {author} {\bibfnamefont {M.}~\bibnamefont {M{\"o}ller}}, \bibinfo {author} {\bibfnamefont {K.~E.}\ \bibnamefont {Priebe}}, \bibinfo {author} {\bibfnamefont {T.}~\bibnamefont {Domr{\"o}se}}, \bibinfo {author} {\bibfnamefont {J.~G.}\ \bibnamefont {Gatzmann}}, \bibinfo {author} {\bibfnamefont {S.}~\bibnamefont {Rost}}, \bibinfo {author} {\bibfnamefont {J.}~\bibnamefont {Schauss}},  \emph {et~al.},\ }\href@noop {} {\bibfield  {journal} {\bibinfo  {journal} {Ultramicroscopy}\ }\textbf {\bibinfo {volume} {176}},\ \bibinfo {pages} {63} (\bibinfo {year} {2017})}\BibitemShut {NoStop}%
\bibitem [{\citenamefont {Houdellier}\ \emph {et~al.}(2018)\citenamefont {Houdellier}, \citenamefont {Caruso}, \citenamefont {Weber}, \citenamefont {Kociak},\ and\ \citenamefont {Arbouet}}]{houdellier2018development}%
  \BibitemOpen
  \bibfield  {author} {\bibinfo {author} {\bibfnamefont {F.}~\bibnamefont {Houdellier}}, \bibinfo {author} {\bibfnamefont {G.~M.}\ \bibnamefont {Caruso}}, \bibinfo {author} {\bibfnamefont {S.}~\bibnamefont {Weber}}, \bibinfo {author} {\bibfnamefont {M.}~\bibnamefont {Kociak}}, \ and\ \bibinfo {author} {\bibfnamefont {A.}~\bibnamefont {Arbouet}},\ }\href@noop {} {\bibfield  {journal} {\bibinfo  {journal} {Ultramicroscopy}\ }\textbf {\bibinfo {volume} {186}},\ \bibinfo {pages} {128} (\bibinfo {year} {2018})}\BibitemShut {NoStop}%
\bibitem [{\citenamefont {Zhu}\ \emph {et~al.}(2020)\citenamefont {Zhu}, \citenamefont {Zheng}, \citenamefont {Wang}, \citenamefont {Zhang}, \citenamefont {Li}, \citenamefont {Sun}, \citenamefont {Xu}, \citenamefont {Tian}, \citenamefont {Li}, \citenamefont {Yang} \emph {et~al.}}]{zhu2020development}%
  \BibitemOpen
  \bibfield  {author} {\bibinfo {author} {\bibfnamefont {C.}~\bibnamefont {Zhu}}, \bibinfo {author} {\bibfnamefont {D.}~\bibnamefont {Zheng}}, \bibinfo {author} {\bibfnamefont {H.}~\bibnamefont {Wang}}, \bibinfo {author} {\bibfnamefont {M.}~\bibnamefont {Zhang}}, \bibinfo {author} {\bibfnamefont {Z.}~\bibnamefont {Li}}, \bibinfo {author} {\bibfnamefont {S.}~\bibnamefont {Sun}}, \bibinfo {author} {\bibfnamefont {P.}~\bibnamefont {Xu}}, \bibinfo {author} {\bibfnamefont {H.}~\bibnamefont {Tian}}, \bibinfo {author} {\bibfnamefont {Z.}~\bibnamefont {Li}}, \bibinfo {author} {\bibfnamefont {H.}~\bibnamefont {Yang}},  \emph {et~al.},\ }\href@noop {} {\bibfield  {journal} {\bibinfo  {journal} {Ultramicroscopy}\ }\textbf {\bibinfo {volume} {209}},\ \bibinfo {pages} {112887} (\bibinfo {year} {2020})}\BibitemShut {NoStop}%
\bibitem [{\citenamefont {Maurer}\ \emph {et~al.}(2011)\citenamefont {Maurer}, \citenamefont {Jesacher}, \citenamefont {Bernet},\ and\ \citenamefont {Ritsch-Marte}}]{maurer2011spatial}%
  \BibitemOpen
  \bibfield  {author} {\bibinfo {author} {\bibfnamefont {C.}~\bibnamefont {Maurer}}, \bibinfo {author} {\bibfnamefont {A.}~\bibnamefont {Jesacher}}, \bibinfo {author} {\bibfnamefont {S.}~\bibnamefont {Bernet}}, \ and\ \bibinfo {author} {\bibfnamefont {M.}~\bibnamefont {Ritsch-Marte}},\ }\href@noop {} {\bibfield  {journal} {\bibinfo  {journal} {Laser \& Photonics Reviews}\ }\textbf {\bibinfo {volume} {5}},\ \bibinfo {pages} {81} (\bibinfo {year} {2011})}\BibitemShut {NoStop}%
\bibitem [{\citenamefont {Grillo}\ \emph {et~al.}(2014)\citenamefont {Grillo}, \citenamefont {Karimi}, \citenamefont {Gazzadi}, \citenamefont {Frabboni}, \citenamefont {Dennis},\ and\ \citenamefont {Boyd}}]{grillo2014}%
  \BibitemOpen
  \bibfield  {author} {\bibinfo {author} {\bibfnamefont {V.}~\bibnamefont {Grillo}}, \bibinfo {author} {\bibfnamefont {E.}~\bibnamefont {Karimi}}, \bibinfo {author} {\bibfnamefont {G.~C.}\ \bibnamefont {Gazzadi}}, \bibinfo {author} {\bibfnamefont {S.}~\bibnamefont {Frabboni}}, \bibinfo {author} {\bibfnamefont {M.~R.}\ \bibnamefont {Dennis}}, \ and\ \bibinfo {author} {\bibfnamefont {R.~W.}\ \bibnamefont {Boyd}},\ }\href {\doibase 10.1103/PhysRevX.4.011013} {\bibfield  {journal} {\bibinfo  {journal} {Phys. Rev. X}\ }\textbf {\bibinfo {volume} {4}},\ \bibinfo {pages} {011013} (\bibinfo {year} {2014})}\BibitemShut {NoStop}%
\bibitem [{\citenamefont {Shiloh}\ \emph {et~al.}(2014)\citenamefont {Shiloh}, \citenamefont {Lereah}, \citenamefont {Lilach},\ and\ \citenamefont {Arie}}]{SHILOH201426}%
  \BibitemOpen
  \bibfield  {author} {\bibinfo {author} {\bibfnamefont {R.}~\bibnamefont {Shiloh}}, \bibinfo {author} {\bibfnamefont {Y.}~\bibnamefont {Lereah}}, \bibinfo {author} {\bibfnamefont {Y.}~\bibnamefont {Lilach}}, \ and\ \bibinfo {author} {\bibfnamefont {A.}~\bibnamefont {Arie}},\ }\href {\doibase https://doi.org/10.1016/j.ultramic.2014.04.007} {\bibfield  {journal} {\bibinfo  {journal} {Ultramicroscopy}\ }\textbf {\bibinfo {volume} {144}},\ \bibinfo {pages} {26} (\bibinfo {year} {2014})}\BibitemShut {NoStop}%
\bibitem [{\citenamefont {Shiloh}\ \emph {et~al.}(2018)\citenamefont {Shiloh}, \citenamefont {Remez}, \citenamefont {Lu}, \citenamefont {Jin}, \citenamefont {Lereah}, \citenamefont {Tavabi}, \citenamefont {Dunin-Borkowski},\ and\ \citenamefont {Arie}}]{shiloh2018spherical}%
  \BibitemOpen
  \bibfield  {author} {\bibinfo {author} {\bibfnamefont {R.}~\bibnamefont {Shiloh}}, \bibinfo {author} {\bibfnamefont {R.}~\bibnamefont {Remez}}, \bibinfo {author} {\bibfnamefont {P.-H.}\ \bibnamefont {Lu}}, \bibinfo {author} {\bibfnamefont {L.}~\bibnamefont {Jin}}, \bibinfo {author} {\bibfnamefont {Y.}~\bibnamefont {Lereah}}, \bibinfo {author} {\bibfnamefont {A.~H.}\ \bibnamefont {Tavabi}}, \bibinfo {author} {\bibfnamefont {R.~E.}\ \bibnamefont {Dunin-Borkowski}}, \ and\ \bibinfo {author} {\bibfnamefont {A.}~\bibnamefont {Arie}},\ }\href {https://www.sciencedirect.com/science/article/pii/S0304399117305259} {\bibfield  {journal} {\bibinfo  {journal} {Ultramicroscopy}\ }\textbf {\bibinfo {volume} {189}},\ \bibinfo {pages} {46} (\bibinfo {year} {2018})}\BibitemShut {NoStop}%
\bibitem [{\citenamefont {Verbeeck}\ \emph {et~al.}(2018)\citenamefont {Verbeeck}, \citenamefont {Béché}, \citenamefont {Müller-Caspary}, \citenamefont {Guzzinati}, \citenamefont {Luong},\ and\ \citenamefont {{Den Hertog}}}]{VERBEECK201858}%
  \BibitemOpen
  \bibfield  {author} {\bibinfo {author} {\bibfnamefont {J.}~\bibnamefont {Verbeeck}}, \bibinfo {author} {\bibfnamefont {A.}~\bibnamefont {Béché}}, \bibinfo {author} {\bibfnamefont {K.}~\bibnamefont {Müller-Caspary}}, \bibinfo {author} {\bibfnamefont {G.}~\bibnamefont {Guzzinati}}, \bibinfo {author} {\bibfnamefont {M.~A.}\ \bibnamefont {Luong}}, \ and\ \bibinfo {author} {\bibfnamefont {M.}~\bibnamefont {{Den Hertog}}},\ }\href {\doibase https://doi.org/10.1016/j.ultramic.2018.03.017} {\bibfield  {journal} {\bibinfo  {journal} {Ultramicroscopy}\ }\textbf {\bibinfo {volume} {190}},\ \bibinfo {pages} {58} (\bibinfo {year} {2018})}\BibitemShut {NoStop}%
\bibitem [{\citenamefont {Ibáñez}\ \emph {et~al.}(2022)\citenamefont {Ibáñez}, \citenamefont {Béché},\ and\ \citenamefont {Verbeeck}}]{ibanez2022}%
  \BibitemOpen
  \bibfield  {author} {\bibinfo {author} {\bibfnamefont {F.~V.}\ \bibnamefont {Ibáñez}}, \bibinfo {author} {\bibfnamefont {A.}~\bibnamefont {Béché}}, \ and\ \bibinfo {author} {\bibfnamefont {J.}~\bibnamefont {Verbeeck}},\ }\href {\doibase 10.48550/ARXIV.2205.07697} {\enquote {\bibinfo {title} {Can a programmable phase plate serve as an aberration corrector in the transmission electron microscope (tem)?}}\ } (\bibinfo {year} {2022})\BibitemShut {NoStop}%
\bibitem [{\citenamefont {Roitman}\ \emph {et~al.}(2021)\citenamefont {Roitman}, \citenamefont {Shiloh}, \citenamefont {Lu}, \citenamefont {Dunin-Borkowski},\ and\ \citenamefont {Arie}}]{roitman2021shaping}%
  \BibitemOpen
  \bibfield  {author} {\bibinfo {author} {\bibfnamefont {D.}~\bibnamefont {Roitman}}, \bibinfo {author} {\bibfnamefont {R.}~\bibnamefont {Shiloh}}, \bibinfo {author} {\bibfnamefont {P.-H.}\ \bibnamefont {Lu}}, \bibinfo {author} {\bibfnamefont {R.~E.}\ \bibnamefont {Dunin-Borkowski}}, \ and\ \bibinfo {author} {\bibfnamefont {A.}~\bibnamefont {Arie}},\ }\href@noop {} {\bibfield  {journal} {\bibinfo  {journal} {ACS photonics}\ }\textbf {\bibinfo {volume} {8}},\ \bibinfo {pages} {3394} (\bibinfo {year} {2021})}\BibitemShut {NoStop}%
\bibitem [{\citenamefont {Ribet}\ \emph {et~al.}(2023)\citenamefont {Ribet}, \citenamefont {Zeltmann}, \citenamefont {Bustillo}, \citenamefont {Dhall}, \citenamefont {Denes}, \citenamefont {Minor}, \citenamefont {Dos~Reis}, \citenamefont {Dravid},\ and\ \citenamefont {Ophus}}]{ribet2023design}%
  \BibitemOpen
  \bibfield  {author} {\bibinfo {author} {\bibfnamefont {S.~M.}\ \bibnamefont {Ribet}}, \bibinfo {author} {\bibfnamefont {S.~E.}\ \bibnamefont {Zeltmann}}, \bibinfo {author} {\bibfnamefont {K.~C.}\ \bibnamefont {Bustillo}}, \bibinfo {author} {\bibfnamefont {R.}~\bibnamefont {Dhall}}, \bibinfo {author} {\bibfnamefont {P.}~\bibnamefont {Denes}}, \bibinfo {author} {\bibfnamefont {A.~M.}\ \bibnamefont {Minor}}, \bibinfo {author} {\bibfnamefont {R.}~\bibnamefont {Dos~Reis}}, \bibinfo {author} {\bibfnamefont {V.~P.}\ \bibnamefont {Dravid}}, \ and\ \bibinfo {author} {\bibfnamefont {C.}~\bibnamefont {Ophus}},\ }\href@noop {} {\bibfield  {journal} {\bibinfo  {journal} {Microscopy and Microanalysis}\ }\textbf {\bibinfo {volume} {29}},\ \bibinfo {pages} {1950} (\bibinfo {year} {2023})}\BibitemShut {NoStop}%
\bibitem [{\citenamefont {Yu}\ \emph {et~al.}(2023)\citenamefont {Yu}, \citenamefont {Vega~Iba{\~n}ez}, \citenamefont {B{\'e}ch{\'e}},\ and\ \citenamefont {Verbeeck}}]{yu2023quantum}%
  \BibitemOpen
  \bibfield  {author} {\bibinfo {author} {\bibfnamefont {C.-P.}\ \bibnamefont {Yu}}, \bibinfo {author} {\bibfnamefont {F.}~\bibnamefont {Vega~Iba{\~n}ez}}, \bibinfo {author} {\bibfnamefont {A.}~\bibnamefont {B{\'e}ch{\'e}}}, \ and\ \bibinfo {author} {\bibfnamefont {J.}~\bibnamefont {Verbeeck}},\ }\href@noop {} {\bibfield  {journal} {\bibinfo  {journal} {SciPost Physics}\ }\textbf {\bibinfo {volume} {15}},\ \bibinfo {pages} {223} (\bibinfo {year} {2023})}\BibitemShut {NoStop}%
\bibitem [{\citenamefont {Barwick}\ \emph {et~al.}(2009)\citenamefont {Barwick}, \citenamefont {Flannigan},\ and\ \citenamefont {Zewail}}]{barwick2009photon}%
  \BibitemOpen
  \bibfield  {author} {\bibinfo {author} {\bibfnamefont {B.}~\bibnamefont {Barwick}}, \bibinfo {author} {\bibfnamefont {D.~J.}\ \bibnamefont {Flannigan}}, \ and\ \bibinfo {author} {\bibfnamefont {A.~H.}\ \bibnamefont {Zewail}},\ }\href {https://www.nature.com/articles/nature08662} {\bibfield  {journal} {\bibinfo  {journal} {Nature}\ }\textbf {\bibinfo {volume} {462}},\ \bibinfo {pages} {902} (\bibinfo {year} {2009})}\BibitemShut {NoStop}%
\bibitem [{\citenamefont {Feist}\ \emph {et~al.}(2015)\citenamefont {Feist}, \citenamefont {Echternkamp}, \citenamefont {Schauss}, \citenamefont {Yalunin}, \citenamefont {Sch{\"a}fer},\ and\ \citenamefont {Ropers}}]{feist2015quantum}%
  \BibitemOpen
  \bibfield  {author} {\bibinfo {author} {\bibfnamefont {A.}~\bibnamefont {Feist}}, \bibinfo {author} {\bibfnamefont {K.~E.}\ \bibnamefont {Echternkamp}}, \bibinfo {author} {\bibfnamefont {J.}~\bibnamefont {Schauss}}, \bibinfo {author} {\bibfnamefont {S.~V.}\ \bibnamefont {Yalunin}}, \bibinfo {author} {\bibfnamefont {S.}~\bibnamefont {Sch{\"a}fer}}, \ and\ \bibinfo {author} {\bibfnamefont {C.}~\bibnamefont {Ropers}},\ }\href@noop {} {\bibfield  {journal} {\bibinfo  {journal} {Nature}\ }\textbf {\bibinfo {volume} {521}},\ \bibinfo {pages} {200} (\bibinfo {year} {2015})}\BibitemShut {NoStop}%
\bibitem [{\citenamefont {Vanacore}\ \emph {et~al.}(2018)\citenamefont {Vanacore}, \citenamefont {Madan}, \citenamefont {Berruto}, \citenamefont {Wang}, \citenamefont {Pomarico}, \citenamefont {Lamb}, \citenamefont {McGrouther}, \citenamefont {Kaminer}, \citenamefont {Barwick}, \citenamefont {Garc{\'\i}a~de Abajo} \emph {et~al.}}]{vanacore2018attosecond}%
  \BibitemOpen
  \bibfield  {author} {\bibinfo {author} {\bibfnamefont {G.~M.}\ \bibnamefont {Vanacore}}, \bibinfo {author} {\bibfnamefont {I.}~\bibnamefont {Madan}}, \bibinfo {author} {\bibfnamefont {G.}~\bibnamefont {Berruto}}, \bibinfo {author} {\bibfnamefont {K.}~\bibnamefont {Wang}}, \bibinfo {author} {\bibfnamefont {E.}~\bibnamefont {Pomarico}}, \bibinfo {author} {\bibfnamefont {R.}~\bibnamefont {Lamb}}, \bibinfo {author} {\bibfnamefont {D.}~\bibnamefont {McGrouther}}, \bibinfo {author} {\bibfnamefont {I.}~\bibnamefont {Kaminer}}, \bibinfo {author} {\bibfnamefont {B.}~\bibnamefont {Barwick}}, \bibinfo {author} {\bibfnamefont {F.~J.}\ \bibnamefont {Garc{\'\i}a~de Abajo}},  \emph {et~al.},\ }\href@noop {} {\bibfield  {journal} {\bibinfo  {journal} {Nature communications}\ }\textbf {\bibinfo {volume} {9}},\ \bibinfo {pages} {2694} (\bibinfo {year} {2018})}\BibitemShut {NoStop}%
\bibitem [{\citenamefont {Vanacore}\ \emph {et~al.}(2019)\citenamefont {Vanacore}, \citenamefont {Berruto}, \citenamefont {Madan}, \citenamefont {Pomarico}, \citenamefont {Biagioni}, \citenamefont {Lamb}, \citenamefont {McGrouther}, \citenamefont {Reinhardt}, \citenamefont {Kaminer}, \citenamefont {Barwick} \emph {et~al.}}]{vanacore2019ultrafast}%
  \BibitemOpen
  \bibfield  {author} {\bibinfo {author} {\bibfnamefont {G.~M.}\ \bibnamefont {Vanacore}}, \bibinfo {author} {\bibfnamefont {G.}~\bibnamefont {Berruto}}, \bibinfo {author} {\bibfnamefont {I.}~\bibnamefont {Madan}}, \bibinfo {author} {\bibfnamefont {E.}~\bibnamefont {Pomarico}}, \bibinfo {author} {\bibfnamefont {P.}~\bibnamefont {Biagioni}}, \bibinfo {author} {\bibfnamefont {R.}~\bibnamefont {Lamb}}, \bibinfo {author} {\bibfnamefont {D.}~\bibnamefont {McGrouther}}, \bibinfo {author} {\bibfnamefont {O.}~\bibnamefont {Reinhardt}}, \bibinfo {author} {\bibfnamefont {I.}~\bibnamefont {Kaminer}}, \bibinfo {author} {\bibfnamefont {B.}~\bibnamefont {Barwick}},  \emph {et~al.},\ }\href@noop {} {\bibfield  {journal} {\bibinfo  {journal} {Nature materials}\ }\textbf {\bibinfo {volume} {18}},\ \bibinfo {pages} {573} (\bibinfo {year} {2019})}\BibitemShut {NoStop}%
\bibitem [{\citenamefont {Kone{\v{c}}n{\'a}}\ and\ \citenamefont {de~Abajo}(2020)}]{konevcna2020electron}%
  \BibitemOpen
  \bibfield  {author} {\bibinfo {author} {\bibfnamefont {A.}~\bibnamefont {Kone{\v{c}}n{\'a}}}\ and\ \bibinfo {author} {\bibfnamefont {F.~J.~G.}\ \bibnamefont {de~Abajo}},\ }\href@noop {} {\bibfield  {journal} {\bibinfo  {journal} {Physical Review Letters}\ }\textbf {\bibinfo {volume} {125}},\ \bibinfo {pages} {030801} (\bibinfo {year} {2020})}\BibitemShut {NoStop}%
\bibitem [{\citenamefont {Ben~Hayun}\ \emph {et~al.}(2021)\citenamefont {Ben~Hayun}, \citenamefont {Reinhardt}, \citenamefont {Nemirovsky}, \citenamefont {Karnieli}, \citenamefont {Rivera},\ and\ \citenamefont {Kaminer}}]{ben2021shaping}%
  \BibitemOpen
  \bibfield  {author} {\bibinfo {author} {\bibfnamefont {A.}~\bibnamefont {Ben~Hayun}}, \bibinfo {author} {\bibfnamefont {O.}~\bibnamefont {Reinhardt}}, \bibinfo {author} {\bibfnamefont {J.}~\bibnamefont {Nemirovsky}}, \bibinfo {author} {\bibfnamefont {A.}~\bibnamefont {Karnieli}}, \bibinfo {author} {\bibfnamefont {N.}~\bibnamefont {Rivera}}, \ and\ \bibinfo {author} {\bibfnamefont {I.}~\bibnamefont {Kaminer}},\ }\href@noop {} {\bibfield  {journal} {\bibinfo  {journal} {Science Advances}\ }\textbf {\bibinfo {volume} {7}},\ \bibinfo {pages} {eabe4270} (\bibinfo {year} {2021})}\BibitemShut {NoStop}%
\bibitem [{\citenamefont {Shiloh}\ \emph {et~al.}(2021)\citenamefont {Shiloh}, \citenamefont {Illmer}, \citenamefont {Chlouba}, \citenamefont {Yousefi}, \citenamefont {Sch{\"o}nenberger}, \citenamefont {Niedermayer}, \citenamefont {Mittelbach},\ and\ \citenamefont {Hommelhoff}}]{shiloh2021electron}%
  \BibitemOpen
  \bibfield  {author} {\bibinfo {author} {\bibfnamefont {R.}~\bibnamefont {Shiloh}}, \bibinfo {author} {\bibfnamefont {J.}~\bibnamefont {Illmer}}, \bibinfo {author} {\bibfnamefont {T.}~\bibnamefont {Chlouba}}, \bibinfo {author} {\bibfnamefont {P.}~\bibnamefont {Yousefi}}, \bibinfo {author} {\bibfnamefont {N.}~\bibnamefont {Sch{\"o}nenberger}}, \bibinfo {author} {\bibfnamefont {U.}~\bibnamefont {Niedermayer}}, \bibinfo {author} {\bibfnamefont {A.}~\bibnamefont {Mittelbach}}, \ and\ \bibinfo {author} {\bibfnamefont {P.}~\bibnamefont {Hommelhoff}},\ }\href@noop {} {\bibfield  {journal} {\bibinfo  {journal} {Nature}\ }\textbf {\bibinfo {volume} {597}},\ \bibinfo {pages} {498} (\bibinfo {year} {2021})}\BibitemShut {NoStop}%
\bibitem [{\citenamefont {Henke}\ \emph {et~al.}(2021)\citenamefont {Henke}, \citenamefont {Raja}, \citenamefont {Feist}, \citenamefont {Huang}, \citenamefont {Arend}, \citenamefont {Yang}, \citenamefont {Kappert}, \citenamefont {Wang}, \citenamefont {M{\"o}ller}, \citenamefont {Pan} \emph {et~al.}}]{henke2021integrated}%
  \BibitemOpen
  \bibfield  {author} {\bibinfo {author} {\bibfnamefont {J.-W.}\ \bibnamefont {Henke}}, \bibinfo {author} {\bibfnamefont {A.~S.}\ \bibnamefont {Raja}}, \bibinfo {author} {\bibfnamefont {A.}~\bibnamefont {Feist}}, \bibinfo {author} {\bibfnamefont {G.}~\bibnamefont {Huang}}, \bibinfo {author} {\bibfnamefont {G.}~\bibnamefont {Arend}}, \bibinfo {author} {\bibfnamefont {Y.}~\bibnamefont {Yang}}, \bibinfo {author} {\bibfnamefont {F.~J.}\ \bibnamefont {Kappert}}, \bibinfo {author} {\bibfnamefont {R.~N.}\ \bibnamefont {Wang}}, \bibinfo {author} {\bibfnamefont {M.}~\bibnamefont {M{\"o}ller}}, \bibinfo {author} {\bibfnamefont {J.}~\bibnamefont {Pan}},  \emph {et~al.},\ }\href@noop {} {\bibfield  {journal} {\bibinfo  {journal} {Nature}\ }\textbf {\bibinfo {volume} {600}},\ \bibinfo {pages} {653} (\bibinfo {year} {2021})}\BibitemShut {NoStop}%
\bibitem [{\citenamefont {Dahan}\ \emph {et~al.}(2021)\citenamefont {Dahan}, \citenamefont {Gorlach}, \citenamefont {Haeusler}, \citenamefont {Karnieli}, \citenamefont {Eyal}, \citenamefont {Yousefi}, \citenamefont {Segev}, \citenamefont {Arie}, \citenamefont {Eisenstein}, \citenamefont {Hommelhoff} \emph {et~al.}}]{dahan2021imprinting}%
  \BibitemOpen
  \bibfield  {author} {\bibinfo {author} {\bibfnamefont {R.}~\bibnamefont {Dahan}}, \bibinfo {author} {\bibfnamefont {A.}~\bibnamefont {Gorlach}}, \bibinfo {author} {\bibfnamefont {U.}~\bibnamefont {Haeusler}}, \bibinfo {author} {\bibfnamefont {A.}~\bibnamefont {Karnieli}}, \bibinfo {author} {\bibfnamefont {O.}~\bibnamefont {Eyal}}, \bibinfo {author} {\bibfnamefont {P.}~\bibnamefont {Yousefi}}, \bibinfo {author} {\bibfnamefont {M.}~\bibnamefont {Segev}}, \bibinfo {author} {\bibfnamefont {A.}~\bibnamefont {Arie}}, \bibinfo {author} {\bibfnamefont {G.}~\bibnamefont {Eisenstein}}, \bibinfo {author} {\bibfnamefont {P.}~\bibnamefont {Hommelhoff}},  \emph {et~al.},\ }\href@noop {} {\bibfield  {journal} {\bibinfo  {journal} {Science}\ }\textbf {\bibinfo {volume} {373}},\ \bibinfo {pages} {eabj7128} (\bibinfo {year} {2021})}\BibitemShut {NoStop}%
\bibitem [{\citenamefont {Feist}\ \emph {et~al.}(2022)\citenamefont {Feist}, \citenamefont {Huang}, \citenamefont {Arend}, \citenamefont {Yang}, \citenamefont {Henke}, \citenamefont {Raja}, \citenamefont {Kappert}, \citenamefont {Wang}, \citenamefont {Louren{\c{c}}o-Martins}, \citenamefont {Qiu} \emph {et~al.}}]{feist2022cavity}%
  \BibitemOpen
  \bibfield  {author} {\bibinfo {author} {\bibfnamefont {A.}~\bibnamefont {Feist}}, \bibinfo {author} {\bibfnamefont {G.}~\bibnamefont {Huang}}, \bibinfo {author} {\bibfnamefont {G.}~\bibnamefont {Arend}}, \bibinfo {author} {\bibfnamefont {Y.}~\bibnamefont {Yang}}, \bibinfo {author} {\bibfnamefont {J.-W.}\ \bibnamefont {Henke}}, \bibinfo {author} {\bibfnamefont {A.~S.}\ \bibnamefont {Raja}}, \bibinfo {author} {\bibfnamefont {F.~J.}\ \bibnamefont {Kappert}}, \bibinfo {author} {\bibfnamefont {R.~N.}\ \bibnamefont {Wang}}, \bibinfo {author} {\bibfnamefont {H.}~\bibnamefont {Louren{\c{c}}o-Martins}}, \bibinfo {author} {\bibfnamefont {Z.}~\bibnamefont {Qiu}},  \emph {et~al.},\ }\href@noop {} {\bibfield  {journal} {\bibinfo  {journal} {Science}\ }\textbf {\bibinfo {volume} {377}},\ \bibinfo {pages} {777} (\bibinfo {year} {2022})}\BibitemShut {NoStop}%
\bibitem [{\citenamefont {Madan}\ \emph {et~al.}(2022)\citenamefont {Madan}, \citenamefont {Leccese}, \citenamefont {Mazur}, \citenamefont {Barantani}, \citenamefont {LaGrange}, \citenamefont {Sapozhnik}, \citenamefont {Tengdin}, \citenamefont {Gargiulo}, \citenamefont {Rotunno}, \citenamefont {Olaya} \emph {et~al.}}]{madan2022ultrafast}%
  \BibitemOpen
  \bibfield  {author} {\bibinfo {author} {\bibfnamefont {I.}~\bibnamefont {Madan}}, \bibinfo {author} {\bibfnamefont {V.}~\bibnamefont {Leccese}}, \bibinfo {author} {\bibfnamefont {A.}~\bibnamefont {Mazur}}, \bibinfo {author} {\bibfnamefont {F.}~\bibnamefont {Barantani}}, \bibinfo {author} {\bibfnamefont {T.}~\bibnamefont {LaGrange}}, \bibinfo {author} {\bibfnamefont {A.}~\bibnamefont {Sapozhnik}}, \bibinfo {author} {\bibfnamefont {P.~M.}\ \bibnamefont {Tengdin}}, \bibinfo {author} {\bibfnamefont {S.}~\bibnamefont {Gargiulo}}, \bibinfo {author} {\bibfnamefont {E.}~\bibnamefont {Rotunno}}, \bibinfo {author} {\bibfnamefont {J.-C.}\ \bibnamefont {Olaya}},  \emph {et~al.},\ }\href@noop {} {\bibfield  {journal} {\bibinfo  {journal} {ACS photonics}\ }\textbf {\bibinfo {volume} {9}},\ \bibinfo {pages} {3215} (\bibinfo {year} {2022})}\BibitemShut {NoStop}%
\bibitem [{\citenamefont {Tsesses}\ \emph {et~al.}(2023)\citenamefont {Tsesses}, \citenamefont {Dahan}, \citenamefont {Wang}, \citenamefont {Bucher}, \citenamefont {Cohen}, \citenamefont {Reinhardt}, \citenamefont {Bartal},\ and\ \citenamefont {Kaminer}}]{tsesses2023tunable}%
  \BibitemOpen
  \bibfield  {author} {\bibinfo {author} {\bibfnamefont {S.}~\bibnamefont {Tsesses}}, \bibinfo {author} {\bibfnamefont {R.}~\bibnamefont {Dahan}}, \bibinfo {author} {\bibfnamefont {K.}~\bibnamefont {Wang}}, \bibinfo {author} {\bibfnamefont {T.}~\bibnamefont {Bucher}}, \bibinfo {author} {\bibfnamefont {K.}~\bibnamefont {Cohen}}, \bibinfo {author} {\bibfnamefont {O.}~\bibnamefont {Reinhardt}}, \bibinfo {author} {\bibfnamefont {G.}~\bibnamefont {Bartal}}, \ and\ \bibinfo {author} {\bibfnamefont {I.}~\bibnamefont {Kaminer}},\ }\href@noop {} {\bibfield  {journal} {\bibinfo  {journal} {Nature Materials}\ }\textbf {\bibinfo {volume} {22}},\ \bibinfo {pages} {345} (\bibinfo {year} {2023})}\BibitemShut {NoStop}%
\bibitem [{\citenamefont {Garc{\'\i}a~de Abajo}\ and\ \citenamefont {Ropers}(2023)}]{garcia2023spatiotemporal}%
  \BibitemOpen
  \bibfield  {author} {\bibinfo {author} {\bibfnamefont {F.~J.}\ \bibnamefont {Garc{\'\i}a~de Abajo}}\ and\ \bibinfo {author} {\bibfnamefont {C.}~\bibnamefont {Ropers}},\ }\href@noop {} {\bibfield  {journal} {\bibinfo  {journal} {Physical Review Letters}\ }\textbf {\bibinfo {volume} {130}},\ \bibinfo {pages} {246901} (\bibinfo {year} {2023})}\BibitemShut {NoStop}%
\bibitem [{\citenamefont {Gaida}\ \emph {et~al.}(2023)\citenamefont {Gaida}, \citenamefont {Louren{\c{c}}o-Martins}, \citenamefont {Yalunin}, \citenamefont {Feist}, \citenamefont {Sivis}, \citenamefont {Hohage}, \citenamefont {Garc{\'\i}a~de Abajo},\ and\ \citenamefont {Ropers}}]{gaida2023lorentz}%
  \BibitemOpen
  \bibfield  {author} {\bibinfo {author} {\bibfnamefont {J.~H.}\ \bibnamefont {Gaida}}, \bibinfo {author} {\bibfnamefont {H.}~\bibnamefont {Louren{\c{c}}o-Martins}}, \bibinfo {author} {\bibfnamefont {S.~V.}\ \bibnamefont {Yalunin}}, \bibinfo {author} {\bibfnamefont {A.}~\bibnamefont {Feist}}, \bibinfo {author} {\bibfnamefont {M.}~\bibnamefont {Sivis}}, \bibinfo {author} {\bibfnamefont {T.}~\bibnamefont {Hohage}}, \bibinfo {author} {\bibfnamefont {F.~J.}\ \bibnamefont {Garc{\'\i}a~de Abajo}}, \ and\ \bibinfo {author} {\bibfnamefont {C.}~\bibnamefont {Ropers}},\ }\href@noop {} {\bibfield  {journal} {\bibinfo  {journal} {Nature Communications}\ }\textbf {\bibinfo {volume} {14}},\ \bibinfo {pages} {6545} (\bibinfo {year} {2023})}\BibitemShut {NoStop}%
\bibitem [{\citenamefont {Synanidis}\ \emph {et~al.}(2024)\citenamefont {Synanidis}, \citenamefont {Gon{\c{c}}alves}, \citenamefont {Ropers},\ and\ \citenamefont {de~Abajo}}]{synanidis2024quantum}%
  \BibitemOpen
  \bibfield  {author} {\bibinfo {author} {\bibfnamefont {A.~P.}\ \bibnamefont {Synanidis}}, \bibinfo {author} {\bibfnamefont {P.}~\bibnamefont {Gon{\c{c}}alves}}, \bibinfo {author} {\bibfnamefont {C.}~\bibnamefont {Ropers}}, \ and\ \bibinfo {author} {\bibfnamefont {F.~J.~G.}\ \bibnamefont {de~Abajo}},\ }\href@noop {} {\bibfield  {journal} {\bibinfo  {journal} {Science Advances}\ }\textbf {\bibinfo {volume} {10}},\ \bibinfo {pages} {eadp4096} (\bibinfo {year} {2024})}\BibitemShut {NoStop}%
\bibitem [{\citenamefont {Fang}\ \emph {et~al.}(2024)\citenamefont {Fang}, \citenamefont {Kuttruff}, \citenamefont {Nabben},\ and\ \citenamefont {Baum}}]{fang2024structured}%
  \BibitemOpen
  \bibfield  {author} {\bibinfo {author} {\bibfnamefont {Y.}~\bibnamefont {Fang}}, \bibinfo {author} {\bibfnamefont {J.}~\bibnamefont {Kuttruff}}, \bibinfo {author} {\bibfnamefont {D.}~\bibnamefont {Nabben}}, \ and\ \bibinfo {author} {\bibfnamefont {P.}~\bibnamefont {Baum}},\ }\href@noop {} {\bibfield  {journal} {\bibinfo  {journal} {Science}\ }\textbf {\bibinfo {volume} {385}},\ \bibinfo {pages} {183} (\bibinfo {year} {2024})}\BibitemShut {NoStop}%
\bibitem [{\citenamefont {Bucksbaum}\ \emph {et~al.}(1988)\citenamefont {Bucksbaum}, \citenamefont {Schumacher},\ and\ \citenamefont {Bashkansky}}]{Bucksbaum1988}%
  \BibitemOpen
  \bibfield  {author} {\bibinfo {author} {\bibfnamefont {P.~H.}\ \bibnamefont {Bucksbaum}}, \bibinfo {author} {\bibfnamefont {D.~W.}\ \bibnamefont {Schumacher}}, \ and\ \bibinfo {author} {\bibfnamefont {M.}~\bibnamefont {Bashkansky}},\ }\href {\doibase 10.1103/PhysRevLett.61.1182} {\bibfield  {journal} {\bibinfo  {journal} {Physical Review Letters}\ }\textbf {\bibinfo {volume} {61}},\ \bibinfo {pages} {1182} (\bibinfo {year} {1988})}\BibitemShut {NoStop}%
\bibitem [{\citenamefont {Freimund}\ \emph {et~al.}(2001)\citenamefont {Freimund}, \citenamefont {Aflatooni},\ and\ \citenamefont {Batelaan}}]{freimund2001observation}%
  \BibitemOpen
  \bibfield  {author} {\bibinfo {author} {\bibfnamefont {D.~L.}\ \bibnamefont {Freimund}}, \bibinfo {author} {\bibfnamefont {K.}~\bibnamefont {Aflatooni}}, \ and\ \bibinfo {author} {\bibfnamefont {H.}~\bibnamefont {Batelaan}},\ }\href@noop {} {\bibfield  {journal} {\bibinfo  {journal} {Nature}\ }\textbf {\bibinfo {volume} {413}},\ \bibinfo {pages} {142} (\bibinfo {year} {2001})}\BibitemShut {NoStop}%
\bibitem [{\citenamefont {Freimund}\ and\ \citenamefont {Batelaan}(2002)}]{freimund2002bragg}%
  \BibitemOpen
  \bibfield  {author} {\bibinfo {author} {\bibfnamefont {D.~L.}\ \bibnamefont {Freimund}}\ and\ \bibinfo {author} {\bibfnamefont {H.}~\bibnamefont {Batelaan}},\ }\href {https://journals.aps.org/prl/abstract/10.1103/PhysRevLett.89.283602} {\bibfield  {journal} {\bibinfo  {journal} {Physical review letters}\ }\textbf {\bibinfo {volume} {89}},\ \bibinfo {pages} {283602} (\bibinfo {year} {2002})}\BibitemShut {NoStop}%
\bibitem [{\citenamefont {Dwyer}\ \emph {et~al.}(2006)\citenamefont {Dwyer}, \citenamefont {Hebeisen}, \citenamefont {Ernstorfer}, \citenamefont {Harb}, \citenamefont {Deyirmenjian}, \citenamefont {Jordan},\ and\ \citenamefont {Dwayne~Miller}}]{dwyer2006femtosecond}%
  \BibitemOpen
  \bibfield  {author} {\bibinfo {author} {\bibfnamefont {J.~R.}\ \bibnamefont {Dwyer}}, \bibinfo {author} {\bibfnamefont {C.~T.}\ \bibnamefont {Hebeisen}}, \bibinfo {author} {\bibfnamefont {R.}~\bibnamefont {Ernstorfer}}, \bibinfo {author} {\bibfnamefont {M.}~\bibnamefont {Harb}}, \bibinfo {author} {\bibfnamefont {V.~B.}\ \bibnamefont {Deyirmenjian}}, \bibinfo {author} {\bibfnamefont {R.~E.}\ \bibnamefont {Jordan}}, \ and\ \bibinfo {author} {\bibfnamefont {R.}~\bibnamefont {Dwayne~Miller}},\ }\href {https://royalsocietypublishing.org/doi/abs/10.1098/rsta.2005.1735} {\bibfield  {journal} {\bibinfo  {journal} {Philosophical Transactions of the Royal Society A: Mathematical, Physical and Engineering Sciences}\ }\textbf {\bibinfo {volume} {364}},\ \bibinfo {pages} {741} (\bibinfo {year} {2006})}\BibitemShut {NoStop}%
\bibitem [{\citenamefont {Baum}\ and\ \citenamefont {Zewail}(2007)}]{baum2007attosecond}%
  \BibitemOpen
  \bibfield  {author} {\bibinfo {author} {\bibfnamefont {P.}~\bibnamefont {Baum}}\ and\ \bibinfo {author} {\bibfnamefont {A.~H.}\ \bibnamefont {Zewail}},\ }\href@noop {} {\bibfield  {journal} {\bibinfo  {journal} {Proceedings of the National Academy of Sciences}\ }\textbf {\bibinfo {volume} {104}},\ \bibinfo {pages} {18409} (\bibinfo {year} {2007})}\BibitemShut {NoStop}%
\bibitem [{\citenamefont {Koz{\'{a}}k}\ \emph {et~al.}(2018)\citenamefont {Koz{\'{a}}k}, \citenamefont {Sch{\"{o}}nenberger},\ and\ \citenamefont {Hommelhoff}}]{Kozak2018}%
  \BibitemOpen
  \bibfield  {author} {\bibinfo {author} {\bibfnamefont {M.}~\bibnamefont {Koz{\'{a}}k}}, \bibinfo {author} {\bibfnamefont {N.}~\bibnamefont {Sch{\"{o}}nenberger}}, \ and\ \bibinfo {author} {\bibfnamefont {P.}~\bibnamefont {Hommelhoff}},\ }\href {\doibase 10.1103/PhysRevLett.120.103203} {\bibfield  {journal} {\bibinfo  {journal} {Physical Review Letters}\ }\textbf {\bibinfo {volume} {120}},\ \bibinfo {pages} {103203} (\bibinfo {year} {2018})}\BibitemShut {NoStop}%
\bibitem [{\citenamefont {Schwartz}\ \emph {et~al.}(2019)\citenamefont {Schwartz}, \citenamefont {Axelrod}, \citenamefont {Campbell}, \citenamefont {Turnbaugh}, \citenamefont {Glaeser},\ and\ \citenamefont {M{\"u}ller}}]{schwartz2019laser}%
  \BibitemOpen
  \bibfield  {author} {\bibinfo {author} {\bibfnamefont {O.}~\bibnamefont {Schwartz}}, \bibinfo {author} {\bibfnamefont {J.~J.}\ \bibnamefont {Axelrod}}, \bibinfo {author} {\bibfnamefont {S.~L.}\ \bibnamefont {Campbell}}, \bibinfo {author} {\bibfnamefont {C.}~\bibnamefont {Turnbaugh}}, \bibinfo {author} {\bibfnamefont {R.~M.}\ \bibnamefont {Glaeser}}, \ and\ \bibinfo {author} {\bibfnamefont {H.}~\bibnamefont {M{\"u}ller}},\ }\href {https://www.nature.com/articles/s41592-019-0552-2} {\bibfield  {journal} {\bibinfo  {journal} {Nature methods}\ }\textbf {\bibinfo {volume} {16}},\ \bibinfo {pages} {1016} (\bibinfo {year} {2019})}\BibitemShut {NoStop}%
\bibitem [{\citenamefont {Axelrod}\ \emph {et~al.}(2020)\citenamefont {Axelrod}, \citenamefont {Campbell}, \citenamefont {Schwartz}, \citenamefont {Turnbaugh}, \citenamefont {Glaeser},\ and\ \citenamefont {M{\"u}ller}}]{axelrod2020observation}%
  \BibitemOpen
  \bibfield  {author} {\bibinfo {author} {\bibfnamefont {J.~J.}\ \bibnamefont {Axelrod}}, \bibinfo {author} {\bibfnamefont {S.~L.}\ \bibnamefont {Campbell}}, \bibinfo {author} {\bibfnamefont {O.}~\bibnamefont {Schwartz}}, \bibinfo {author} {\bibfnamefont {C.}~\bibnamefont {Turnbaugh}}, \bibinfo {author} {\bibfnamefont {R.~M.}\ \bibnamefont {Glaeser}}, \ and\ \bibinfo {author} {\bibfnamefont {H.}~\bibnamefont {M{\"u}ller}},\ }\href@noop {} {\bibfield  {journal} {\bibinfo  {journal} {Physical review letters}\ }\textbf {\bibinfo {volume} {124}},\ \bibinfo {pages} {174801} (\bibinfo {year} {2020})}\BibitemShut {NoStop}%
\bibitem [{\citenamefont {Chirita~Mihaila}\ \emph {et~al.}(2022)\citenamefont {Chirita~Mihaila}, \citenamefont {Weber}, \citenamefont {Schneller}, \citenamefont {Grandits}, \citenamefont {Nimmrichter},\ and\ \citenamefont {Juffmann}}]{chirita2022transverse}%
  \BibitemOpen
  \bibfield  {author} {\bibinfo {author} {\bibfnamefont {M.~C.}\ \bibnamefont {Chirita~Mihaila}}, \bibinfo {author} {\bibfnamefont {P.}~\bibnamefont {Weber}}, \bibinfo {author} {\bibfnamefont {M.}~\bibnamefont {Schneller}}, \bibinfo {author} {\bibfnamefont {L.}~\bibnamefont {Grandits}}, \bibinfo {author} {\bibfnamefont {S.}~\bibnamefont {Nimmrichter}}, \ and\ \bibinfo {author} {\bibfnamefont {T.}~\bibnamefont {Juffmann}},\ }\href@noop {} {\bibfield  {journal} {\bibinfo  {journal} {Physical Review X}\ }\textbf {\bibinfo {volume} {12}},\ \bibinfo {pages} {031043} (\bibinfo {year} {2022})}\BibitemShut {NoStop}%
\bibitem [{\citenamefont {Tsarev}\ \emph {et~al.}(2023)\citenamefont {Tsarev}, \citenamefont {Thurner},\ and\ \citenamefont {Baum}}]{tsarev2023nonlinear}%
  \BibitemOpen
  \bibfield  {author} {\bibinfo {author} {\bibfnamefont {M.}~\bibnamefont {Tsarev}}, \bibinfo {author} {\bibfnamefont {J.~W.}\ \bibnamefont {Thurner}}, \ and\ \bibinfo {author} {\bibfnamefont {P.}~\bibnamefont {Baum}},\ }\href@noop {} {\bibfield  {journal} {\bibinfo  {journal} {Nature Physics}\ }\textbf {\bibinfo {volume} {19}},\ \bibinfo {pages} {1350} (\bibinfo {year} {2023})}\BibitemShut {NoStop}%
\bibitem [{\citenamefont {Ebel}\ and\ \citenamefont {Talebi}(2023)}]{ebel2023inelastic}%
  \BibitemOpen
  \bibfield  {author} {\bibinfo {author} {\bibfnamefont {S.}~\bibnamefont {Ebel}}\ and\ \bibinfo {author} {\bibfnamefont {N.}~\bibnamefont {Talebi}},\ }\href@noop {} {\bibfield  {journal} {\bibinfo  {journal} {Communications Physics}\ }\textbf {\bibinfo {volume} {6}},\ \bibinfo {pages} {179} (\bibinfo {year} {2023})}\BibitemShut {NoStop}%
\bibitem [{\citenamefont {Schaap}\ \emph {et~al.}(2023)\citenamefont {Schaap}, \citenamefont {Sweers}, \citenamefont {Smorenburg},\ and\ \citenamefont {Luiten}}]{schaap2023ponderomotive}%
  \BibitemOpen
  \bibfield  {author} {\bibinfo {author} {\bibfnamefont {B.}~\bibnamefont {Schaap}}, \bibinfo {author} {\bibfnamefont {C.}~\bibnamefont {Sweers}}, \bibinfo {author} {\bibfnamefont {P.}~\bibnamefont {Smorenburg}}, \ and\ \bibinfo {author} {\bibfnamefont {O.}~\bibnamefont {Luiten}},\ }\href@noop {} {\bibfield  {journal} {\bibinfo  {journal} {Physical Review Accelerators and Beams}\ }\textbf {\bibinfo {volume} {26}},\ \bibinfo {pages} {074401} (\bibinfo {year} {2023})}\BibitemShut {NoStop}%
\bibitem [{\citenamefont {Velasco}\ and\ \citenamefont {de~Abajo}(2024)}]{velasco2024free}%
  \BibitemOpen
  \bibfield  {author} {\bibinfo {author} {\bibfnamefont {C.~I.}\ \bibnamefont {Velasco}}\ and\ \bibinfo {author} {\bibfnamefont {F.}~\bibnamefont {de~Abajo}},\ }\href {https://arxiv.org/abs/2412.03410} {\bibfield  {journal} {\bibinfo  {journal} {arXiv preprint}\ }\textbf {\bibinfo {volume} {arXiv:2412.03410}} (\bibinfo {year} {2024})}\BibitemShut {NoStop}%
\bibitem [{\citenamefont {de~Abajo}\ and\ \citenamefont {Kone{\v{c}}n{\'a}}(2021)}]{de2021optical}%
  \BibitemOpen
  \bibfield  {author} {\bibinfo {author} {\bibfnamefont {F.~J.~G.}\ \bibnamefont {de~Abajo}}\ and\ \bibinfo {author} {\bibfnamefont {A.}~\bibnamefont {Kone{\v{c}}n{\'a}}},\ }\href@noop {} {\bibfield  {journal} {\bibinfo  {journal} {Physical Review Letters}\ }\textbf {\bibinfo {volume} {126}},\ \bibinfo {pages} {123901} (\bibinfo {year} {2021})}\BibitemShut {NoStop}%
\bibitem [{\citenamefont {Uesugi}\ \emph {et~al.}(2021)\citenamefont {Uesugi}, \citenamefont {Kozawa},\ and\ \citenamefont {Sato}}]{uesugi2021electron}%
  \BibitemOpen
  \bibfield  {author} {\bibinfo {author} {\bibfnamefont {Y.}~\bibnamefont {Uesugi}}, \bibinfo {author} {\bibfnamefont {Y.}~\bibnamefont {Kozawa}}, \ and\ \bibinfo {author} {\bibfnamefont {S.}~\bibnamefont {Sato}},\ }\href@noop {} {\bibfield  {journal} {\bibinfo  {journal} {Physical Review Applied}\ }\textbf {\bibinfo {volume} {16}},\ \bibinfo {pages} {L011002} (\bibinfo {year} {2021})}\BibitemShut {NoStop}%
\bibitem [{\citenamefont {Uesugi}\ \emph {et~al.}(2022)\citenamefont {Uesugi}, \citenamefont {Kozawa},\ and\ \citenamefont {Sato}}]{uesugi2022properties}%
  \BibitemOpen
  \bibfield  {author} {\bibinfo {author} {\bibfnamefont {Y.}~\bibnamefont {Uesugi}}, \bibinfo {author} {\bibfnamefont {Y.}~\bibnamefont {Kozawa}}, \ and\ \bibinfo {author} {\bibfnamefont {S.}~\bibnamefont {Sato}},\ }\href@noop {} {\bibfield  {journal} {\bibinfo  {journal} {Journal of Optics}\ }\textbf {\bibinfo {volume} {24}},\ \bibinfo {pages} {054013} (\bibinfo {year} {2022})}\BibitemShut {NoStop}%
\bibitem [{\citenamefont {Mihaila}\ and\ \citenamefont {Koz\'{a}k}(2025)}]{ChiritaMihaila:25}%
  \BibitemOpen
  \bibfield  {author} {\bibinfo {author} {\bibfnamefont {M.~C.~C.}\ \bibnamefont {Mihaila}}\ and\ \bibinfo {author} {\bibfnamefont {M.}~\bibnamefont {Koz\'{a}k}},\ }\href {\doibase 10.1364/OE.542930} {\bibfield  {journal} {\bibinfo  {journal} {Opt. Express}\ }\textbf {\bibinfo {volume} {33}},\ \bibinfo {pages} {758} (\bibinfo {year} {2025})}\BibitemShut {NoStop}%
\bibitem [{\citenamefont {Nekula}\ \emph {et~al.}(2025)\citenamefont {Nekula}, \citenamefont {Juffmann},\ and\ \citenamefont {Kone{\v{c}}n{\'a}}}]{nekula2025laser}%
  \BibitemOpen
  \bibfield  {author} {\bibinfo {author} {\bibfnamefont {Z.}~\bibnamefont {Nekula}}, \bibinfo {author} {\bibfnamefont {T.}~\bibnamefont {Juffmann}}, \ and\ \bibinfo {author} {\bibfnamefont {A.}~\bibnamefont {Kone{\v{c}}n{\'a}}},\ }\href@noop {} {\bibfield  {journal} {\bibinfo  {journal} {arXiv preprint arXiv:2501.16501}\ } (\bibinfo {year} {2025})}\BibitemShut {NoStop}%
\bibitem [{\citenamefont {Haider}\ \emph {et~al.}(2000)\citenamefont {Haider}, \citenamefont {Uhlemann},\ and\ \citenamefont {Zach}}]{haider2000upper}%
  \BibitemOpen
  \bibfield  {author} {\bibinfo {author} {\bibfnamefont {M.}~\bibnamefont {Haider}}, \bibinfo {author} {\bibfnamefont {S.}~\bibnamefont {Uhlemann}}, \ and\ \bibinfo {author} {\bibfnamefont {J.}~\bibnamefont {Zach}},\ }\href@noop {} {\bibfield  {journal} {\bibinfo  {journal} {Ultramicroscopy}\ }\textbf {\bibinfo {volume} {81}},\ \bibinfo {pages} {163} (\bibinfo {year} {2000})}\BibitemShut {NoStop}%
\bibitem [{\citenamefont {Kibble}(1966)}]{kibble1966mutual}%
  \BibitemOpen
  \bibfield  {author} {\bibinfo {author} {\bibfnamefont {T.}~\bibnamefont {Kibble}},\ }\href@noop {} {\bibfield  {journal} {\bibinfo  {journal} {Physical Review}\ }\textbf {\bibinfo {volume} {150}},\ \bibinfo {pages} {1060} (\bibinfo {year} {1966})}\BibitemShut {NoStop}%
\bibitem [{\citenamefont {Hu}\ \emph {et~al.}(2020)\citenamefont {Hu}, \citenamefont {Wang}, \citenamefont {Wang}, \citenamefont {Ji}, \citenamefont {Zhang}, \citenamefont {Jiawen}, \citenamefont {Zhu}, \citenamefont {Wu},\ and\ \citenamefont {Chu}}]{Yanlei2020}%
  \BibitemOpen
  \bibfield  {author} {\bibinfo {author} {\bibfnamefont {Y.}~\bibnamefont {Hu}}, \bibinfo {author} {\bibfnamefont {Z.}~\bibnamefont {Wang}}, \bibinfo {author} {\bibfnamefont {X.}~\bibnamefont {Wang}}, \bibinfo {author} {\bibfnamefont {S.}~\bibnamefont {Ji}}, \bibinfo {author} {\bibfnamefont {C.}~\bibnamefont {Zhang}}, \bibinfo {author} {\bibfnamefont {L.}~\bibnamefont {Jiawen}}, \bibinfo {author} {\bibfnamefont {W.}~\bibnamefont {Zhu}}, \bibinfo {author} {\bibfnamefont {D.}~\bibnamefont {Wu}}, \ and\ \bibinfo {author} {\bibfnamefont {J.}~\bibnamefont {Chu}},\ }\href {\doibase 10.1038/s41377-020-00362-z} {\bibfield  {journal} {\bibinfo  {journal} {Light Sci Appl}\ }\textbf {\bibinfo {volume} {9}} (\bibinfo {year} {2020}),\ 10.1038/s41377-020-00362-z}\BibitemShut {NoStop}%
\bibitem [{\citenamefont {Lakshminarayanan}\ and\ \citenamefont {Fleck}(2011)}]{lakshminarayanan2011zernike}%
  \BibitemOpen
  \bibfield  {author} {\bibinfo {author} {\bibfnamefont {V.}~\bibnamefont {Lakshminarayanan}}\ and\ \bibinfo {author} {\bibfnamefont {A.}~\bibnamefont {Fleck}},\ }\href@noop {} {\bibfield  {journal} {\bibinfo  {journal} {Journal of Modern Optics}\ }\textbf {\bibinfo {volume} {58}},\ \bibinfo {pages} {545} (\bibinfo {year} {2011})}\BibitemShut {NoStop}%
\bibitem [{\citenamefont {Kimoto}(2014)}]{kimoto2014practical}%
  \BibitemOpen
  \bibfield  {author} {\bibinfo {author} {\bibfnamefont {K.}~\bibnamefont {Kimoto}},\ }\href@noop {} {\bibfield  {journal} {\bibinfo  {journal} {Journal of Electron Microscopy}\ }\textbf {\bibinfo {volume} {63}},\ \bibinfo {pages} {337} (\bibinfo {year} {2014})}\BibitemShut {NoStop}%
\bibitem [{\citenamefont {Streshkova}\ \emph {et~al.}(2024)\citenamefont {Streshkova}, \citenamefont {Koutensk{\`y}}, \citenamefont {Novotn{\`y}},\ and\ \citenamefont {Koz{\'a}k}}]{streshkova2024monochromatization}%
  \BibitemOpen
  \bibfield  {author} {\bibinfo {author} {\bibfnamefont {N.~L.}\ \bibnamefont {Streshkova}}, \bibinfo {author} {\bibfnamefont {P.}~\bibnamefont {Koutensk{\`y}}}, \bibinfo {author} {\bibfnamefont {T.}~\bibnamefont {Novotn{\`y}}}, \ and\ \bibinfo {author} {\bibfnamefont {M.}~\bibnamefont {Koz{\'a}k}},\ }\href@noop {} {\bibfield  {journal} {\bibinfo  {journal} {Physical Review Letters}\ }\textbf {\bibinfo {volume} {133}},\ \bibinfo {pages} {213801} (\bibinfo {year} {2024})}\BibitemShut {NoStop}%
\bibitem [{\citenamefont {Xie}\ \emph {et~al.}(2015)\citenamefont {Xie}, \citenamefont {Ren}, \citenamefont {Huang}, \citenamefont {Lavery}, \citenamefont {Ahmed}, \citenamefont {Yan}, \citenamefont {Bao}, \citenamefont {Li}, \citenamefont {Zhao}, \citenamefont {Cao} \emph {et~al.}}]{xie2015phase}%
  \BibitemOpen
  \bibfield  {author} {\bibinfo {author} {\bibfnamefont {G.}~\bibnamefont {Xie}}, \bibinfo {author} {\bibfnamefont {Y.}~\bibnamefont {Ren}}, \bibinfo {author} {\bibfnamefont {H.}~\bibnamefont {Huang}}, \bibinfo {author} {\bibfnamefont {M.~P.}\ \bibnamefont {Lavery}}, \bibinfo {author} {\bibfnamefont {N.}~\bibnamefont {Ahmed}}, \bibinfo {author} {\bibfnamefont {Y.}~\bibnamefont {Yan}}, \bibinfo {author} {\bibfnamefont {C.}~\bibnamefont {Bao}}, \bibinfo {author} {\bibfnamefont {L.}~\bibnamefont {Li}}, \bibinfo {author} {\bibfnamefont {Z.}~\bibnamefont {Zhao}}, \bibinfo {author} {\bibfnamefont {Y.}~\bibnamefont {Cao}},  \emph {et~al.},\ }\href@noop {} {\bibfield  {journal} {\bibinfo  {journal} {Optics letters}\ }\textbf {\bibinfo {volume} {40}},\ \bibinfo {pages} {1197} (\bibinfo {year} {2015})}\BibitemShut {NoStop}%
\bibitem [{\citenamefont {Peccianti}\ \emph {et~al.}(2012)\citenamefont {Peccianti}, \citenamefont {Pasquazi}, \citenamefont {Park}, \citenamefont {Little}, \citenamefont {Chu}, \citenamefont {Moss},\ and\ \citenamefont {Morandotti}}]{peccianti2012demonstration}%
  \BibitemOpen
  \bibfield  {author} {\bibinfo {author} {\bibfnamefont {M.}~\bibnamefont {Peccianti}}, \bibinfo {author} {\bibfnamefont {A.}~\bibnamefont {Pasquazi}}, \bibinfo {author} {\bibfnamefont {Y.}~\bibnamefont {Park}}, \bibinfo {author} {\bibfnamefont {B.~E.}\ \bibnamefont {Little}}, \bibinfo {author} {\bibfnamefont {S.~T.}\ \bibnamefont {Chu}}, \bibinfo {author} {\bibfnamefont {D.~J.}\ \bibnamefont {Moss}}, \ and\ \bibinfo {author} {\bibfnamefont {R.}~\bibnamefont {Morandotti}},\ }\href@noop {} {\bibfield  {journal} {\bibinfo  {journal} {Nature communications}\ }\textbf {\bibinfo {volume} {3}},\ \bibinfo {pages} {765} (\bibinfo {year} {2012})}\BibitemShut {NoStop}%
\bibitem [{\citenamefont {Morimoto}\ and\ \citenamefont {Madsen}(2024)}]{morimoto2024scattering}%
  \BibitemOpen
  \bibfield  {author} {\bibinfo {author} {\bibfnamefont {Y.}~\bibnamefont {Morimoto}}\ and\ \bibinfo {author} {\bibfnamefont {L.~B.}\ \bibnamefont {Madsen}},\ }\href@noop {} {\bibfield  {journal} {\bibinfo  {journal} {New Journal of Physics}\ }\textbf {\bibinfo {volume} {26}},\ \bibinfo {pages} {053012} (\bibinfo {year} {2024})}\BibitemShut {NoStop}%
\bibitem [{\citenamefont {Mihaila}\ \emph {et~al.}(2025)\citenamefont {Mihaila}, \citenamefont {Streshkova},\ and\ \citenamefont {Koz{\'a}k}}]{zenodo_dataset}%
  \BibitemOpen
  \bibfield  {author} {\bibinfo {author} {\bibfnamefont {M.~C.~C.}\ \bibnamefont {Mihaila}}, \bibinfo {author} {\bibfnamefont {N.~L.}\ \bibnamefont {Streshkova}}, \ and\ \bibinfo {author} {\bibfnamefont {M.}~\bibnamefont {Koz{\'a}k}},\ }\href {\doibase 10.5281/zenodo.15298830} {\enquote {\bibinfo {title} {Dataset for "light-based chromatic aberration correction of ultrafast electron microscopes"},}\ } (\bibinfo {year} {2025})\BibitemShut {NoStop}%
\end{thebibliography}%

\clearpage
\onecolumngrid
\setcounter{section}{0}
\setcounter{equation}{0}

\renewcommand{\theequation}{S\arabic{equation}}
\section*{Supplemental Material}

\title{Supplemental Material: Light-based Chromatic Aberration Correction of Ultrafast Electron Microscopes}

\author{Marius Constantin Chirita Mihaila}
  \affiliation{Charles University, Faculty of Mathematics and Physics, Ke Karlovu 3, 121 16 Prague 2}

 \author{Neli Laštovičková Streshkova}
  \affiliation{Charles University, Faculty of Mathematics and Physics, Ke Karlovu 3, 121 16 Prague 2} 

\author{Martin Koz{\'a}k}
  \affiliation{Charles University, Faculty of Mathematics and Physics, Ke Karlovu 3, 121 16 Prague 2}

\date{\today}
             
\maketitle

\section{Ponderomotive potential shaping and calculation of the optical field}
The critical point of the proposed method is the shaping of a suitable optical beam focus profile, indicated by the complex function $\mathbf{\tilde{g}}(\mathbf{r})$. The two simplest options are to use a Gaussian beam (divergent lens) with angular momentum $l=0$ or a Laguerre-Gaussian mode (convergent lens) with orbital angular momentum $l=1$ focused near the interaction plane. Near the beam center, the intensity profile is approximately parabolic in the radial direction, and in a suitable position near the optical focus, the ponderomotive phase changes nearly linearly with $E$. The gradient in the optical field intensity stemming only from the focusing itself can lead to some degree of correction.

State-of-the-art optical shaping devices such as SLMs allow to imprint tailored phase and amplitude modulation on an incoming plane wave. This can be used to shift the focal spot but also to manipulate the optical beam focus to the required shape. In the proposed scheme (see Fig 1. (b) in the main manuscript), the collimated optical beam is incident on the SLM (coordinates $x_r,\,y_r$) with a Gaussian transverse distribution, a flat phase and a polarization vector $\mathbf{e}_{x_{r}}$ along the $x_r$ axis. The SLM imprints a custom phase profile to the initial electric field before propagation:
\begin{align}
    \textbf{E}_i (x_r,y_r) = & c_{0} \exp{[-(x_r^2+y_r^2)/w_{o}^{2}]}  e^{il\phi_r} \mathbf{e}_{x_{r}}\notag \\
    &\exp\Big[ i \pi (c^{0}_{2}Z^{0}_2(x_r,y_r) + c^{0}_{4}Z^{0}_{4}(x_r,y_r) + ... )\Big],
\end{align}
where $c_{0}$ is the electric field amplitude, $\phi_r=\arctan{y/x}$ is the azimuthal angle, $l$ is the topological charge and $Z^{0}_{n}(x_r,y_r)$ denotes the Zernike polynomials~\cite{lakshminarayanan2011zernike} with axial symmetry with weights $c^{0}_{n}$ in the linear combination.

The optical field $\mathbf{E}_i$ is focused with a moderate NA lens with focal distance $f_o$ to achieve a sufficient gradient in the optical intensity around the focus. Following on~\cite{Yanlei2020}, the vectorial Debye diffraction integral, simplified into a Fourier transform form, is expressed as:
\begin{equation}
\mathbf{\Tilde{g}}(x, y, -z) = -i \exp(i\omega z/c) \frac{C}{\lambda_o} \mathcal{F}_{2D} \left[ \mathbf{M} \cdot \mathbf{E}_i \cdot \frac{\exp(-i \omega z \cos \theta/c)}{\sqrt{\cos \theta}} \right],
\label{Vectorial diffraction FFT}
\end{equation}
where we add the multiplication factor $\exp[i\omega z/c]$ to cancel out the quick oscillations so $\tilde{\mathbf{g}}$ is just the envelope function, $\lambda_o$ is the optical wavelength, $\mathbf{M}$ is the polarization transformation matrix accounting for conversion from the entrance pupil to the spherical reference surface,  $\theta$ is the polar angle in the spherical coordinate system, with the light focus as origin, $\cos \theta$ is the angular cosine term required for energy conservation, and $C$ is the apodization factor ensuring proper amplitude scaling. The sign flip of 
$z$ accounts for the fact that the light propagates in the direction opposite to the electron beam.  The polarization transformation matrix is defined as:
\begin{equation}
    \mathbf{M} = 
    \begin{bmatrix}
1 + (\cos{\theta}-1)\cos^2{\theta}& (\cos{\theta} - 1)\cos{\phi_r} \sin{\phi_r} & -\sin{\theta}\cos{\phi_r} \\
(\cos{\theta} - 1)\cos{\phi_r} \sin{\phi}_r & 1 + (\cos{\theta}-1)\sin^2{\theta} & -\sin{\theta}\sin{\phi_r} \\
\sin{\theta}\cos{\phi_r} & \sin{\theta}\sin{\phi_r} & \cos{\theta}
\end{bmatrix}.
\end{equation}
The computations are performed via the Bluestein algorithm~\cite{Yanlei2020}, which is a generalization of the fast Fourier transform algorithm, allowing to compute the intensity profile for an arbitrary region and with the required resolution in a computationally efficient manner.

We can write the complex focus profile Cartesian components as a product of amplitude and phase:
\begin{equation}
    \tilde{g}_{x,y,z}(x,y,-z) = |\tilde{g}_{x,y,z}(x,y,-z)|\exp{[i\phi_{x,y,z}(x,y,-z)]},
\end{equation}
where $|\tilde{g}_{x,y,z}|$ stands for the amplitude of the component, and the $\phi_{x,y,z}$ is the phase correction added to the plane wave propagation phase $\omega z/c$. The complex spatio-temporal representation of the optical pulse in focus is:
\begin{equation}
    \varepsilon_{x,y,z}(\mathbf{r},t) = \exp\left[-\frac{(t - t_0 + \frac{z}{c})^2}{2\sigma_t^2}\right] |\tilde{g}_{x,y,z}(x,y,-z)| e^{i\phi_{x,y,z} (x,y,-z)} e^{-i \omega ( t + z/c)},
\end{equation}
and the vector potential $\mathbf{A}(\mathbf{r}, t) = \Re{\Big[\frac{\bm{\varepsilon}(\mathbf{r}, t)}{i\omega}\Big]}$ written by components is:
\begin{equation}
    A_{x,y,z}(\textbf{r},t)= \frac{1}{\omega}\exp\left[-\frac{(t - t_0 + \frac{z}{c})^2}{2\sigma_t^2}\right] |\tilde{g}_{x,y,z}(x,y,-z)| \sin{\Big[\omega ( t + z/c) - \phi_{x,y,z} (x,y,-z)\Big]}.
\end{equation}
Considering that the interaction takes place near the optical axis and within $\approx 20\,\rm{\mu m}$ in the $z$ direction, we can neglect the phase correction $\phi_{x,y,z}(x,y,-z)$.

The optimal set of weights $c_0$ and $c^0_n$ is found by a gradient descent algorithm~\cite{xie2015phase}, aiming to minimize the dispersion of the sum of the phases $\chi + \varphi$ and the second average derivative $\partial^2_y [\chi + \varphi]$ (see~\ref{Gradient Decent} chapter for details).

\section{Gradient descent algorithm}\label{Gradient Decent}

We implement the gradient descent algorithm to optimize the weights of the Zernike polynomials $c^{0}_{n}$, and the optical field amplitude $c_{0}$. The goal is to shape the ponderomotive potential $\varphi$ in such a way that the focal distance corresponding to the total phase $\chi(y,E) + \varphi(y,E;c_{0}, c^{0}_{n})$ does not depend on the energy within the energy spread of the electron beam.

This condition can be mathematically quantified by the statistical variance $\sigma\{\partial^2_y(\chi+\varphi)\}$ of the second derivative sum $\frac{\partial^{2}}{\partial y^{2}}[\chi(y,E) + \varphi(y,E;c_0,c_n^0)]$, relative to the statistical variance $\sigma\{\frac{\partial ^{2}}{\partial y^2}\chi\}$ of the aberration phase without optical compensation. The statistics is accounted for only within the electron spot $y\leq w_e$, and averaged over all electron energies:
\begin{equation}
    \text{cost} (c_0,c_n^{0}) =\frac{\sigma^{2}\{\frac{\partial^{2}}{\partial y^{2}}(\chi+\varphi)\}}{ \sigma^{2}\{\frac{\partial^{2}}{\partial y^{2}} \chi\}}.
\end{equation}
The aim is to minimize this ratio. The cost function changes smoothly with the parameter variation, and provides reasonable values for the optimized coefficients. The aberration correction causes all the energy components to focus in one plane, which can be shifted from the original focus of the component $E_0$. The required electron lens defocus $\Delta z_0$ needs to be found separatelly. Alternatively, the electron lens defocus can be included in the optimized parameters in an adjusted cost function
\begin{equation}
    \text{cost}_2  (\Delta z_0, c_0,c_n^{0})= \frac{ \sigma^{2}\{\frac{\partial^{2}}{\partial y^{2}} [\chi (y, E;\Delta z_0)+ \varphi(y,E;c_0, c_n^0)]\}}{ \sigma^{2}\{\frac{\partial^{2}}{\partial y^{2}} [\chi (y, E;\Delta z_0 = 0)]\}}.
\end{equation}

To automatize the parameter optimization process, we use $N$ restarts with a randomized choice of input values for $c_0, c_n^{0}$ within given intervals. This procedure also helps to avoid local minima of the cost function. Further, we implement the accumulation of momentum, adaptive learning rate proportional to the magnitude of the parameters, and an early stopping condition, if the gradient descent appears to stagnate and the cost function is not improved sufficiently over several iterations. The optimized values are subsequently applied to compute the corrected electron beam profile, which is then compared to the aberrated profile and the target profile. The coefficients used are $c_2^0 = -1.12,\, c_4^0 = -1.78, \, c_6^{0} = 0.30, \, c_8^{0} = 1.29 $ for the vortex mode (Fig.~2(b) in main manuscript) and $c_2^0 = -0.02,\, c_4^0 = -0.57, \, c_6^{0} = -0.55, \, c_8^{0} = 1.11$ for the Gaussian mode (Fig.~2(c) in main manuscript). The required peak electric field in focus is $27\, \rm{GV/m}$ for vortex-like and $13\, \rm{GV/m}$ for Gaussian-like beams. The required electron lens defocus is $\Delta z_0 = 8.5 \, \rm{\mu m}$ for vortex-like and $\Delta z_0 = -2.5 \,\rm{\mu m}$ for Gaussian-like beams.

\section{Extension Beyond Transverse Phase Shaping
}

While both our previous study~\cite{ChiritaMihaila:25} and this work utilize phase-modulated optical beams to correct aberrations in ultrafast electron microscopes, they target fundamentally different aberrations, spherical versus chromatic, and do so using distinct methodologies.  Spherical aberration results from the over-focusing of peripheral rays at large angles, whereas chromatic aberration stems from focal length variations for electrons with different energies, which are present due to the finite energy spread of the electron beam.

A key advancement in the present work is the introduction of phase modulation that varies not only transversely but also longitudinally along the electron propagation direction. Our earlier study focused solely on transverse shaping of the optical field in one plane to correct spherical aberration and did not consider variations along the longitudinal (z) axis. In contrast, this work explicitly incorporates energy-dependent effects, taking into account how different energy components of a chirped electron pulse interact with the ponderomotive field at different longitudinal positions. This dynamic interaction enables an effective correction of chromatic aberration, which arises from energy-dependent focal shifts, a capability not present in the earlier approach.

Importantly, the ability to correct chromatic aberration has significant implications. It enables ultrafast electron microscopes to maintain high spatial resolution even when operating with broadened electron pulses (e.g., due to the Boersch effect), which would otherwise degrade image quality via energy-dependent focal shifts.

\end{document}